\documentclass[11pt]{article}
\pdfoutput=1
\usepackage{draft}
\usepackage{comment}
\usepackage[normalem]{ulem}
\usepackage{hyperref}
\hypersetup{
 colorlinks=true,
 urlcolor=blue,
 anchorcolor=blue,
 citecolor=blue,
 filecolor=blue,
 linkcolor=blue,
 menucolor=blue,
 pagecolor=blue,
 linktocpage=true,
}
\usepackage{graphicx,color,subfig}
\usepackage{cite}
\usepackage{empheq}

\usepackage{bbold}
\usepackage[usenames,dvipsnames,table]{xcolor}
\usepackage[labelfont=bf]{caption}
\graphicspath{{./figures/}}
\usepackage{tikz}

\newcommand{\RR}{\mathbb{R}}

\newcommand{\fusion}{\mathbb{F}}	
\newcommand{\id}{\mathbb{1}}	

\newcommand{\sbmatrix}[1]{
{\tiny\arraycolsep=0.3\arraycolsep\ensuremath{\begin{bmatrix}#1\end{bmatrix}}}
}

\begin{document}

\unitlength = .8mm

 \begin{titlepage}
 \begin{center}

 \hfill \\
 \hfill \\

\title{Universal Dynamics in Non-Orientable CFT$_2$}

 \author{ Ioannis Tsiares$^{b}$}

 \address{
 $^{b}$Department of Physics, McGill University,
 Montreal, QC H3A 2T8, Canada
 }

 \email{ioannis.tsiares@mail.mcgill.ca}

 \end{center}

 \abstract{
Two-dimensional conformal field theories (CFTs) defined on non-orientable Riemann surfaces obey consistency Cardy conditions analogous to those in the orientable case. We revisit those conditions for irrational theories with central charge $c>1$ in the context of two-point functions of primaries on the Real Projective plane $\mathbb{RP}^2$ and the partition function on the Klein bottle $\mathbb{K}^2$. Using the irrational versions of the Virasoro fusion and modular kernels we derive universal expressions for the non-orientable CFT data at large conformal dimension, assuming a gap in the spectrum of scalar primaries. In particular, we derive asymptotic formulas at finite central charge for the averaged Light-Light-Heavy product $C_{LLH}\times\Gamma_{H}$ of OPE coefficients with the $\mathbb{RP}^2$ one-point function normalizations, as well as for the parity-weighted density of heavy scalar primaries (or equivalently the density of heavy $\Gamma_H^2$). We discuss the gravitational interpretation of the results.
 }

 \vfill

 \end{titlepage}

\eject

\begingroup
\hypersetup{linkcolor=black}

\renewcommand{\baselinestretch}{0.93}\normalsize
\tableofcontents
\renewcommand{\baselinestretch}{1.0}\normalsize

\endgroup

\section{Introduction}\label{sec:intro}
Two dimensional conformal field theories (CFTs) constitute a very special class of quantum field theories which are highly constrained by symmetry. At the minimum, unitarity and the infinite dimensional Virasoro algebra of local conformal transformations are already powerful enough to impose non-trivial constraints in the theory \cite{Belavin:1984vu}. On top of that, Euclidean CFT correlation functions also depend on the conformal structure moduli of the two-dimensional Riemann surface $\Sigma$ on which they are defined, and therefore they should additionally be invariant under different parametrizations of this moduli. This requirement introduces new and a priori distinct constraints on the theory. A prototypical incarnation of this statement is modular invariance of the torus partition function, one important aspect of which leads for example to the celebrated Cardy formula for the asymptotics of the high energy density of states\cite{Cardy:1986ie} (or its large spin version \cite{Kusuki:2018wpa,Kusuki:2019gjs,Benjamin:2019stq,Maxfield:2019hdt}). Other examples involve crossing symmetry of four-point functions on the sphere $S^2$ where one can obtain asymptotics for the OPE spectral density\cite{Das:2017cnv,Collier:2018exn,Collier:2019weq}, modular covariance of one and two-point functions on the torus\cite{Kraus:2016nwo,Brehm:2018ipf,Hikida:2018khg,Romero-Bermudez:2018dim,Collier:2019weq}, or modular invariance of the genus 2 partition function\cite{Cardy:2017qhl,Collier:2019weq}\footnote{The asymptotic formulas that arise in orientable surfaces with Euler character $\chi=2-2g-n_b$ equal to $-2$,  i.e.: i) four-point function on $S^2$, ii) two-point function on the torus and iii) genus two partition function have recently been studied in a unified way in \cite{Collier:2019weq}, where one can see that a single formula (namely, the vacuum fusion kernel) governs the various asymptotics of the OPE-squared spectral density.}.
\par In this paper, we will focus on similar type of constraints that arise in the case of non-orientable surfaces. CFTs on two dimensional non-orientable surfaces have been studied in the past mostly in the context of string theories on non-orientable worldsheets (see e.g. \cite{Blumenhagen:2009zz,Fioravanti:1993hf,Pradisi:1995qy,Pradisi:1995pp}) or under the general framework of two-dimensional boundary CFTs after the work of Cardy, Lewellen\cite{Cardy:1989ir,Cardy:1991tv,Lewellen:1991tb} and Ishibashi\cite{Ishibashi:1988kg} (see e.g. \cite{Blumenhagen:2009zz} for a review). Most of these constructions however involve Rational CFTs with or without extended symmetry algebras and - with the exception of Liouville theory\cite{Hikida:2002bt,Nakayama:2004vk} - far less is known about the generic features of an irrational CFT  on a non-orientable surface with $c>1$ and just Virasoro symmetry. We will make a first step towards filling this gap in this work. Apart from being interesting in its own merit, the formal study of irrational theories is also crucial from a holographic perspective where we expect a weakly coupled theory of gravity in AdS$_3$ with non-orientable boundary to capture (under certain assumptions) the dynamics of a large $c$ CFT living on this boundary\cite{Maldacena:1997re}. Interesting recent work in this direction includes \cite{Verlinde:2015qfa,Nakayama:2015mva,Nakayama:2016xvw, Maloney:2016gsg,Lewkowycz:2016ukf,LeFloch:2017lbt,Wang:2020jgh,Giombi:2020xah,Hogervorst:2017kbj}. 
\par We will consider generic irrational CFTs with $c_{R}=c_{L}=c$ which possess only Virasoro symmetry and have a gap in the spectrum of scalar primaries above the identity. Using consistency conditions for the two-point function of primaries on the Real Projective plane $\mathbb{RP}^2$ and the partition function on the Klein bottle $\mathbb{K}^2$, we will derive asymptotic formulas that \textit{universally} govern the non-orientable CFT data in particular kinematic limits. Our method parallels the derivation of similar asymptotic formulas in the orientable case\cite{Collier:2018exn,Collier:2019weq}, where we implement knowledge of the corresponding vacuum fusion kernel in the appropriate channel. To summarize our results, from the two-point function of identical primaries $O$ on $\mathbb{RP}^2$ we derive\footnote{Throughout this paper we use the notation $a\sim b$ to denote that ${a\over b}\to 1$ in the limit of interest. We will also use the notation $a\approx b$ to denote that $a$ and $b$ have the same leading scaling in the limit of interest.}:
\begin{equation}\label{LLHH}
\begin{aligned}
\overline{C_{OOO_p}\Gamma_{O_p}}&\approx 4^{-\Delta_p}e^{-3\pi\sqrt{\frac{c-1}{12}\Delta_p}}\Delta_p^{2\Delta_O-\frac{c+1}{8}} \ , \ \ \ \ \Delta_p>>c,\Delta_O
\end{aligned}
\end{equation}
for the averaged product of OPE coefficient of the two operators $O$ with a heavy \textit{scalar} primary $O_p$ times the $\mathbb{RP}^2$ one-point function normalization $\Gamma_{O_p}$ of the scalar. From the duality of Klein bottle partition function we obtain:
\begin{equation}\label{H}
\begin{aligned}
\rho_{sc.}^{\pm}(\Delta)&\approx e^{2\pi\sqrt{\frac{c-1}{12}\Delta}}\\
\overline{\Gamma^2_{\Delta}}&\approx e^{-2\pi \sqrt{\frac{c-1}{12}\Delta}} \ , \ \ \ \  \ \ \ \Delta>>c.
\end{aligned}
\end{equation}
where $\rho_{sc.}^{\pm}(\Delta)$ is the parity-weighted spectral density of scalar primaries. We will discuss those derivations in detail in sections \ref{sec:RP2} and \ref{sec:K2} respectively. 
\par On a different ground, the conformal bootstrap program of \cite{Rychkov:2016iqz,Poland:2018epd} has been recently explored in the Real Projective space in higher dimensions \cite{Nakayama:2016cim,Hasegawa:2016piv,Hogervorst:2017kbj,Hasegawa:2018yqg} and  proven a fruitful exercise to obtain new results; nevertheless, far less is known in the two-dimensional case where one can implement the full power of the infinite dimensional Virasoro algebra. With that in mind, our side goal here will also be to initiate a bootstrap problem for the Klein bottle and highlight its importance.
\par The paper is organized as follows: in section \ref{sec:RP2} we review the one and two-point functions on $\mathbb{RP}^2$. Using the crossing symmetry of two-point functions along with the fusion kernel of Ponsot and Teschner, we derive the universal asymptotic expression for the Light-Light-Heavy data $C_{LLH}\times\Gamma_H$ in (\ref{LLHH}). In section \ref{sec:K2} we review the duality of the Klein bottle partition function and in a similar fashion, using the modular kernel, we derive the asymptotic expression for the parity-weighted density of heavy scalars or, equivalently, the density of heavy $\Gamma^2_{H}$ in (\ref{H}). We also formulate the bootstrap problem on the Klein bottle. In section \ref{sec:GR} we analyze the large central charge limits of our asymptotic formulas and discuss their interpretation from the point of view of a weakly-coupled theory of gravity in $AdS_3$ with non-orientable boundary. In appendix \ref{appA} we work out the example of the Klein bottle partition function of the compactified free boson on a radius $R$. Finally, in appendix \ref{appB} we verify the Klein bottle bootstrap equation introduced in section \ref{sec:K2} for the case of the Ising and the tri-critical Ising model.

\section{CFT on $\mathbb{RP}^2$}\label{sec:RP2}

\subsection{Review of the basics}\label{sec:RP2basics}

The \textit{Real Projective plane} or the \textit{Crosscap surface} is a non-orientable two dimensional surface with no boundary and Euler character\footnote{The Euler character of a Riemann surface with $n_b$ boundaries, $n_c$ crosscaps, and $g$ handles is $\chi=2-2g-n_c-n_b$.}$\chi_{\mathbb{RP}^2}=1$. It can be obtained from the disk by pairwise identifying opposite points on its boundary, or from the Riemann sphere $\hat{\mathbb{C}}$ by a fixed point-free antiholomorphic involution $\mathcal{I}$ that changes the orientation:
\begin{figure}
	\centering
	\includegraphics[width=.3\textwidth]{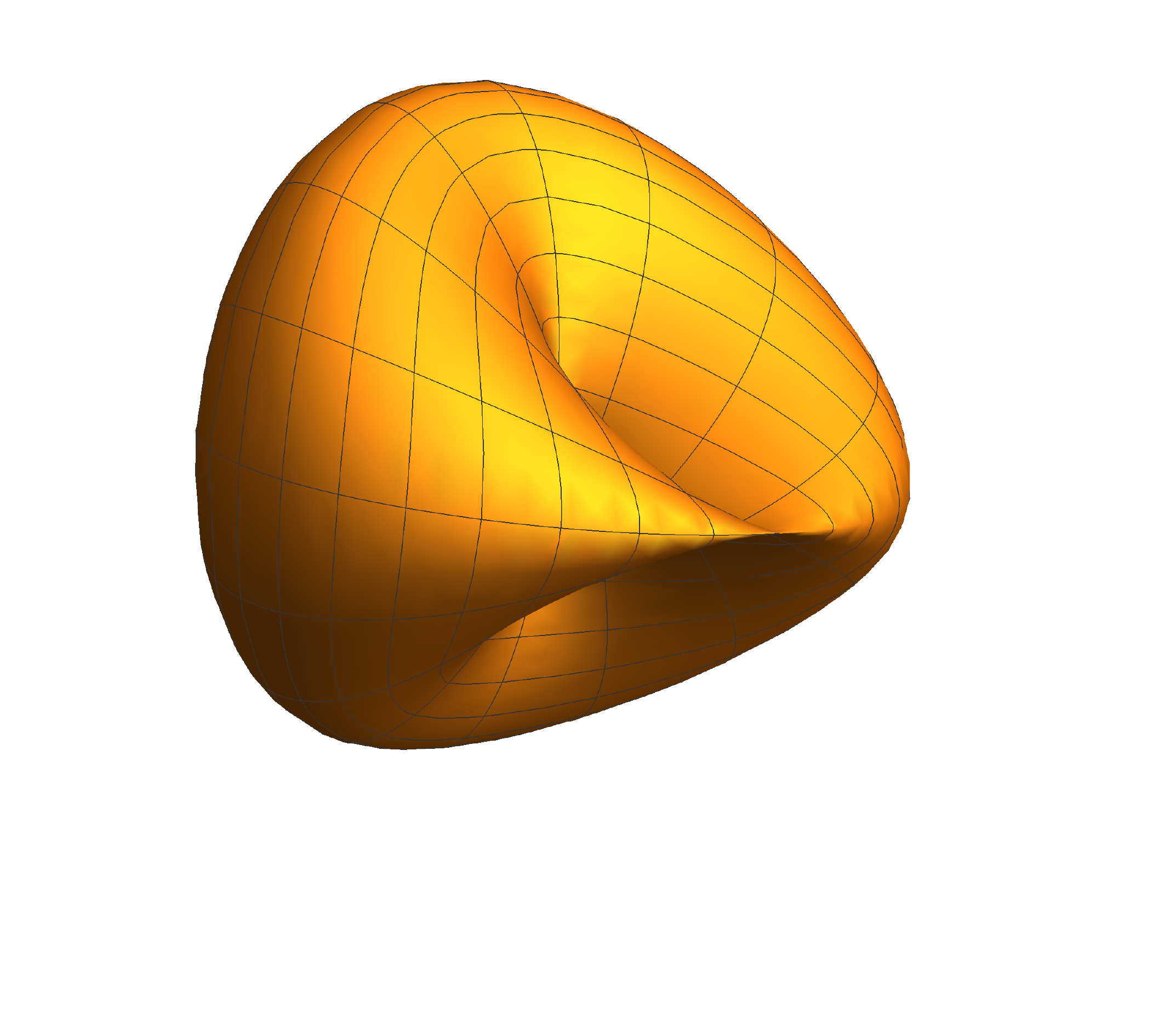} 
\caption[RP2.]{The Crosscap surface.
	\label{fig:KBex}}
	\end{figure}
\begin{equation}\label{invo}
\mathcal{I}: \ z\sim z'=-\frac{1}{\bar{z}}
\end{equation}
where $z,\bar{z}=z^*$ are coordinates on $\hat{\mathbb{C}}$. The fundamental region of the involution is the unit disc. Under (\ref{invo}), a Virasoro primary field $O_i$ of conformal dimensions $(h_i,\overline{h_i})$ at one point is related to its image point as
\begin{equation}\label{invoprim}
\begin{aligned}
\mathcal{I}:O_i(z,\bar{z})&\sim\epsilon_i z^{-2h_i}\overline{z}^{-2\bar{h}_i}O_i\left(-\frac{1}{\bar{z}},-\frac{1}{z}\right)
\end{aligned}
\end{equation}
where $\epsilon_i=\pm 1$ is the eigenvalue of the corresponding primary under the involution (since $\mathcal{I}^2=\mathbb{1}$). Notice that the image field of $O_i$ has essentially \textit{exchanged} conformal dimensions $(\overline{h_i},h_i)$ compared to the original field (we can see that e.g. from the Laurent mode expansion of $O_i\left(-\frac{1}{\bar{z}},-\frac{1}{z}\right)$). From now on we will adopt the Liouville parametrization of the conformal dimensions and central charge as
\begin{equation}\label{}
\begin{aligned}
c&=1+6Q^2=1+6(b+b^{-1})^2>1\\
h_i&=\left(\frac{Q}{2}\right)^2+P_i^2=\alpha_i(Q-\alpha_i) \ ,  \ \ \ \ \ \alpha=\frac{Q}{2}+iP
\end{aligned}
\end{equation}
and similarly for $\bar{P}_i$ or $\bar{\alpha}_i$ in place of $\overline{h_i}$\footnote{The energy eigenvalue on the plane is given by the total scaling dimension $\Delta_i=h_i+\bar{h}_i$.}. We will be using the various parameterizations interchangeably. Note that the Liouville parametrization for the conformal dimensions is invariant under reflections $P\rightarrow-P$ (or $\alpha\rightarrow Q-\alpha$), and it also naturally splits unitary values of the weights ($h\geq0$) into two distinct regimes: i) the \textit{``continuous regime''} $h\geq \frac{c-1}{24}$ corresponding to real $P$ (or $\alpha\in \frac{Q}{2}+i\mathbb{R}$) and ii) the \textit{``discrete regime''} $0\leq h < \frac{c-1}{24}$ corresponding to imaginary $P$ (or $\alpha\in[0,\frac{Q}{2})$).
\par Absence of energy flux at the `boundary' of the unit disc under the involution (\ref{invo}) imposes a conformal boundary condition for the energy-momentum tensor:
\begin{equation}\label{}
\begin{aligned}
T(z)-z^{-4}\overline{T}(\bar{z}=-1/z)=0
\end{aligned}
\end{equation}
which in turn implies that the Virasoro modes should be related (classically) as:
\begin{equation}\label{}
\begin{aligned}
L_n-(-1)^n\overline{L}_{-n}=0 \ , \ \ \ \ \ \ \ \ \ \ \ \  n\in\mathbb{Z} 
\end{aligned}
\end{equation}
Quantum mechanically, we can define a state $\ket{C}$ in the CFT Hilbert space on the circle by
\begin{equation}\label{crosseq}
\begin{aligned}
\left(L_n-(-1)^n\overline{L}_{-n}\right)\ket{C}=0 \ , \ \ \ \ \ \ n\in\mathbb{Z} 
\end{aligned}
\end{equation}
This is the crosscap state, and it's basically the state prepared by the Euclidean CFT path integral on a crosscap surface. From the above expression we can see immediately that $\ket{C}$ is a state of zero spin, though \textit{not} an energy eigenstate. In \cite{Ishibashi:1988kg} it was shown that, for rational CFTs, there is a basis of solutions to (\ref{crosseq}) whose cardinality is equal to the number of \textit{scalar} primaries in the theory. These are the so-called \textit{crosscap Ishibashi states} $| C,P \rangle \rangle$ which satisfy
\begin{equation}\label{ishi}
\begin{aligned}
&\left(L_n-(-1)^n\overline{L}_{-n}\right)| C,P \rangle \rangle=0 \ , \ \ \ \ \ \ n\in\mathbb{Z}
\end{aligned}
\end{equation}
and therefore we can write
\begin{equation}\label{ishi}
\begin{aligned}
&\ket{C}=\sum_{P\in\{\text{scalars}\}}\Gamma_{P}| C,P \rangle \rangle
\end{aligned}
\end{equation}
The sum runs over scalar primaries $P$, and the (real) coefficients $\Gamma_P$ are often called $\mathbb{RP}^2$  \textit{one-point function normalizations}. These coefficients determine the one-point functions of primaries in the presence of a crosscap, as we will see next, and they are non-zero only for scalar primaries with positive eigenvalue $\epsilon_P=+1$. The cross-cap Ishibashi
states can be realized as infinite sums of products of holomorphic and anti-holomorphic states of the form:
\begin{equation}\label{ishibst}
\begin{aligned}
| C,P \rangle \rangle=\sum_{\vec{m}}(-1)^{\sum_{j}m_j}\ket{P,\vec{m}}\otimes U\ket{\overline{P,\vec{m}}}
\end{aligned}
\end{equation}
where $\vec{m}$ denotes the descendant state constructed by acting with $L_{-j}$ raising operator
$m_j$ times on the primary $\ket{P}$, and $U$ is an anti-unitary operator \cite{Ishibashi:1988kg,Blumenhagen:2009zz}. They are also normalized as follows:
\begin{equation}\label{normishi}
\begin{aligned}
\langle\langle C, P' | e^{-a\left(L_0+\overline{L_0}-\frac{c}{12}\right)}| C,P \rangle \rangle&=\delta(P-P')\frac{e^{-2a P^2}}{\eta(\frac{ia}{\pi})}\ , \ \ \ \ P\neq vac. \\
\langle\langle C, vac | e^{-a\left(L_0+\overline{L_0}-\frac{c}{12}\right)}|C, vac \rangle \rangle&=\frac{(1-e^{-2a})e^{\frac{aQ^2}{2}}}{\eta(\frac{ia}{\pi})}
\end{aligned}
\end{equation}
where $\eta(\tau)$ is the Dedekind eta function and $a$ some real constant (we will see later that $a$ is going to play the role of the inverse temperature).
\par In the rest of the paper, we're going to assume that a decomposition of the form (\ref{ishi}) exists and converges for the crosscap state of any irrational CFT with $c>1$ and a given spectrum of scalar primaries\footnote{Although originally the decomposition (\ref{ishi}) was shown to exist in rational theories\cite{Ishibashi:1988kg}, extending this to the irrational case ($c>1$) does not seem to be a harmful statement, since one can construct the Ishibashi state of a given primary as a coherent state level by level in the descendants as in (\ref{ishibst}) (see also e.g. \cite{Blumenhagen:2009zz,Nakayama:2004vk}). In the case of Liouville theory for example the crosscap state has been constructed explicitly at any central charge in terms of Ishibashi states \cite{Hikida:2002bt,Nakayama:2004vk}.} (which is a subset of the general spectrum of the theory). It will also be important for us later to assume a scalar gap above the identity. This would mean that
\begin{equation}\label{}
\begin{aligned}
\ket{C}=\Gamma_{\mathbb{1}}| C,vac \rangle \rangle+\sum_{P_{gap}\neq\mathbb{1}}\Gamma_{P}| C,P \rangle \rangle
\end{aligned}
\end{equation}
where $P_{gap}$ can be either in the discrete or the continuous regime.

\subsection{One-point functions}\label{sec:RP2onept}

We can express the correlation function of a string of operators $\{X\}$ on the Real Projective plane as the amplitude:
\begin{equation}\label{}
\begin{aligned}
\langle X\rangle_{\mathbb{RP}^2}=\bra{0}X\ket{C}
\end{aligned}
\end{equation}
Taking into account (\ref{invoprim}) along with the fact that $\ket{C}$ has zero spin, we can see immediately that only one-point functions of \textit{scalar} primaries $O_i$ with $\epsilon_i=+1$ can have a non-zero result. More specifically, we have
\begin{equation}\label{cor1}
\begin{aligned}
 \bra{0}\mathbb{1}\ket{C}&=\Gamma_{\mathbb{1}}\equiv Z_{\mathbb{RP}^2}\\
\bra{0}O_i(z,\bar{z})\ket{C}&=\frac{\Gamma_i}{(1+z\bar{z})^{2h_i}}
\end{aligned}
\end{equation}
The first amplitude in (\ref{cor1}) is the partition function on the crosscap surface since it's the expectation value of the identity operator $\mathbb{1}$. The second expression is the one-point function of a generic \textit{scalar} operator $O_i$ of dimension $h_i=\overline{h_i}$ and $\epsilon_i=+1$, and its form was first obtained in \cite{Fioravanti:1993hf} using conformal invariance. The coefficients $\Gamma_i$ are exactly the $\mathbb{RP}^2$  one-point function normalizations that entered in the decomposition (\ref{ishi}). We can write equivalently
\begin{equation}\label{corr1}
\begin{aligned}
\Gamma_i=\bra{0}O_i(0)\ket{C}
\end{aligned}
\end{equation}
\par As it turns out, the coefficients $\Gamma_i$ are the only additional data - along with the OPE coefficients and the scalar primary dimensions - that we need in order to construct any CFT correlation function on a non-orientable surface\cite{Fioravanti:1993hf}. Intuitively this can be understood from the fact that any non-orientable surface $\Sigma$ can be constructed as a connected sum\footnote{The connected sum $\Sigma\#\mathbb{RP}^2$ of a surface $\Sigma$ with $\mathbb{RP}^2$ is the procedure of cutting out a disc on $\Sigma$ and gluing in a crosscap.} of $n_c$ copies of $\mathbb{RP}^2$, i.e. $\Sigma=\underbrace{\mathbb{RP}^2\#\mathbb{RP}^2\#\cdots\#\mathbb{RP}^2}_{n_c}$. As we will see next, the coefficients $\Gamma_i$ are in general highly constrained by consistency conditions of correlation functions on such surfaces. In this work, we will be mainly interested in the constraints that arise in two-point functions on $\mathbb{RP}^2$ and the partiton function on $\mathbb{K}^2$.

\subsection{Two-point functions and the Crosscap Constraint}\label{sec:RP2twopt}

\par Let's consider the presence of two primary (but not necessarily scalar) operators $O_1(z,\overline{z})$,$O_2(w,\overline{w})$ of dimensions $(h_1,\overline{h}_1)$,$(h_2,\overline{h}_2)$ on the crosscap:
\begin{equation}\label{tpt}
\begin{aligned}
\langle O_1(z,\overline{z})O_2(w,\overline{w})\rangle_{\mathbb{RP}^2}=\bra{0}O_1(z,\overline{z})O_2(w,\overline{w})\ket{C}
\end{aligned}
\end{equation}
This amplitude is non-zero only if the eigenvalues $\epsilon_1,\epsilon_2$ of the two operators under the involution (\ref{invo}) are equal: as in the case of the one-point function, this is because we want the correlator to be invariant under (\ref{invo}). Let's call the common eigenvalue $\epsilon_1=\epsilon_2=\epsilon_i(=\pm1)$.
\par As first discussed in \cite{Fioravanti:1993hf} (see also \cite{Pradisi:1995qy,Pradisi:1995pp}), the moduli space of the twice-punctured $\mathbb{RP}^2$ is described by a \textit{one real dimensional} cross-ratio:
\begin{equation}\label{}
\eta=\frac{|z-w|^2}{\left(1+|z|^2\right)\left(1+|w|^2\right)}
\end{equation}
which is manifestly invariant under the involution. Therefore the two point function (\ref{tpt}) is - up to a factor that depends on the external dimensions - equal to a function of $\eta$. We can compute this function in two equivalent ways: first, we can take the OPE between $O_1$ and $O_2$ ($\eta\rightarrow0$) which reduces the computation to a sum of one-point functions of (scalar) primaries along with all their descendants on the crosscap:
\begin{equation}\label{}
\begin{aligned}
O_1(z)O_2(w)\sim\sum_{s}C_{12s}(z-w)^{h_s-h_1-h_2}(\bar{z}-\bar{w})^{\overline{h_s}-\overline{h_1}-\overline{h_2}}\mathcal{C}(z-w,\partial)O_s(w)
\end{aligned}
\end{equation}
where $C_{12s}$ are the sphere OPE coefficients and $\mathcal{C}$ is a differential operator encoding the contributions from the descendants of $O_s$ \footnote{We suppressed the dependence on the barred variable for the operators $O_1,O_2,O_s$ and the differential operator $\mathcal{C}$ for clarity. }. Alternatively, we can take the OPE between $O_1$ and the image $\mathcal{I}(O_2)$ ($\eta\rightarrow1$) which reduces the computation again to a sum of one-point functions of scalar primaries and their descendants in a \textit{rotated} crosscap. Using the definition of the image (\ref{invoprim}) this gives an OPE of the form:
\begin{equation}\label{}
\begin{aligned}
O_1(z)\mathcal{I}(O_2)(-1/\bar{w})\sim\epsilon_i\sum_{t}C_{12t}\left(\frac{1+z\bar{w}}{\bar{w}}\right)^{h_t-h_1-\overline{h_2}}\left(\frac{1+\bar{z}w}{w}\right)^{\overline{h_t}-\overline{h_1}-h_2}\mathcal{C}(\partial)O_t\left(-1/\bar{w}\right)
\end{aligned}
\end{equation}
At the end, the two equivalent expansions for the amplitude yield a crossing equation of the form \cite{Fioravanti:1993hf}:
\begin{equation}\label{creq}
\begin{aligned}
\sum_{O_s}C_{12s}\Gamma_s\mathcal{F}\sbmatrix{P_2 & P_1 \\ \bar{P}_1 & \bar{P}_2}(P_s|\eta)=\epsilon_i(-1)^{h_1-\overline{h_1}+h_2-\overline{h_2}}\sum_{O_t}C_{12t}\Gamma_t\mathcal{F}\sbmatrix{P_2 & \bar{P}_1 \\ P_1 & \bar{P}_2}(P_t|1-\eta)
\end{aligned}
\end{equation}
where the phase factor $(-1)^{h_1-\overline{h_1}+h_2-\overline{h_2}}$ comes from the rotated crosscap frame on the second channel, and $\mathcal{F}\sbmatrix{P_2 & P_1 \\ \bar{P}_1 & \bar{P}_2}(P_i|z)$ (similarly $\mathcal{F}\sbmatrix{P_2 & \bar{P}_1 \\ P_1 & \bar{P}_2}(P_t|1-\eta)$) is a \textit{single} copy of the Virasoro block, namely the holomorphic sphere four-point block with \textit{real} cross-ratio $\eta$ and external dimensions $h_1,h_2,\overline{h_1},\overline{h_2}$. This particular configuration of the external dimensions basically comes from the fact that the two point function on the crosscap can be equivalently thought of as the four-point function on the sphere, where the two additional operators are the \textit{images} of $O_1,O_2$ with \textit{exchanged} conformal dimensions ($\overline{h_1},h_1$), ($\overline{h_2},h_2$) respectively. Note also that only \textit{scalar} primaries with \textit{positive} eigenvalue $\epsilon_s=\epsilon_t=+1$ can propagate as internal operators in both channels, since the corresponding $\Gamma_s,\Gamma_t$ vanish otherwise.
\par In \cite{Fioravanti:1993hf,Pradisi:1995qy,Pradisi:1995pp} the authors studied (\ref{creq}) in the case of rational theories with finite number of primaries and, using the \textit{fusion matrices} that relate different channel conformal blocks, they were able to rewrite (\ref{creq}) as a constraint on the CFT data which they called the \textit{crosscap constraint}. We will now apply the same logic here, except we will implement the power of the \textit{fusion kernel} constructed by Ponsot and Teschner \cite{Ponsot:1999uf,Ponsot:2000mt,Teschner:2001rv} to make a statement about irrational CFTs with central charge $c>1$.
\par The defining  relation of the fusion kernel $\fusion_{P_sP_t}\sbmatrix{P_2 & P_1 \\ P_3 & P_4}$ is
\begin{equation}\label{eq:fusionTransformation}
	\mathcal{F}\sbmatrix{P_2 & P_3 \\ P_1 & P_4}(P_t|1-z) = \int_{C} \frac{d P_s}{2}\fusion_{P_sP_t}\sbmatrix{P_2 & P_1 \\ P_3 & P_4}\mathcal{F}\sbmatrix{P_2 & P_1 \\ P_3 & P_4}(P_s|z), \ \ \ \ \ \ \ \ z\in\mathbb{C}-{\{0,1\}}
\end{equation}
in other words it expresses holomorphic Virasoro blocks on the T-channel as a linear combination of S-channel blocks and $z\in\mathbb{C}-{\{0,1\}}$ is the usual sphere four-point cross-ratio. The kernel $\fusion_{P_sP_t}\sbmatrix{P_2 & P_1 \\ P_3 & P_4}$ is a meromorphic function of $Ps,P_t$ and the support $C$ of the integral depends on the external operator unitary dimensions: if $Re\left(\alpha_1+\alpha_2\right)>\frac{Q}{2}$ the contour $C$ can be chosen to run along the whole real line $\mathbb{R}$, whereas if $\alpha_1+\alpha_2<\frac{Q}{2}$ ($\alpha_i$ necessarily real in the discrete regime) some poles of $\fusion_{P_sP_t}$ may cross the contour $C=\mathbb{R}$ and hence the integral acquires additional contributions from the residues of these poles. For a nice exposition of the analytic properties of the fusion kernel see \cite{Collier:2018exn}. The remarkable construction of this kernel by Ponsot and Teschner has found many applications and  attracted a renewed interest in plenty of recent works\cite{Jackson:2014nla,Chang:2015qfa,Chang:2016ftb,Esterlis:2016psv,Mertens:2017mtv,He:2017lrg,Kusuki:2018wpa,Collier:2018exn,Maxfield:2019hdt,Collier:2019weq, Ghosh:2019rcj}. 
\par Adapting to our set up, we consider
\begin{equation}\label{eq:fusionTransformation}
	\mathcal{F}\sbmatrix{P_2 & \bar{P}_1 \\ P_1 & \bar{P}_2}(P_t|1-\eta) = \int_C \frac{d P_s}{2}\fusion_{P_sP_t}\sbmatrix{P_2 & P_1 \\ \bar{P}_1 & \bar{P}_2}\mathcal{F}\sbmatrix{P_2 & P_1 \\ \bar{P}_1 & \bar{P}_2}(P_s|\eta) \ , \ \ \ \eta\in\mathbb{R}
\end{equation}
Using this expression into the crossing equation (\ref{creq}), a trivial manipulation yields a non-trivial constraint on the non-orientable CFT data:
\begin{equation}\label{crossconstr}
\begin{aligned}
\epsilon_i(-1)^{h_1-\overline{h_1}+h_2-\overline{h_2}}c_{12}(P_s)=\int_C\frac{dP_t}{2}\mathbb{F}_{P_sP_t}\sbmatrix{P_2 & P_1 \\ \bar{P}_1 & \bar{P}_2}c_{12}(P_t)
\end{aligned}
\end{equation}
where we defined the spectral density that encodes the non-orientable data for scalar primaries as the \textit{even distribution}\footnote{In the spirit of \cite{Maxfield:2019hdt,Collier:2019weq}, to arrive in (\ref{crossconstr}) we assumed that the distribution (\ref{2ptdensity}) is defined by its integral against all holomorphic Virasoro blocks in (\ref{creq}), i.e. we assumed that the holomorphic Virasoro blocks are \textit{complete} in the relevant space of test functions.}
\begin{equation}\label{2ptdensity}
\begin{aligned}
c_{12}(P_i)&\equiv \sum_{j\in\{\text{scalars}\}}C_{12j}\Gamma_j\left[\delta(P_i-P_j)+\delta(P_i+P_j)\right]
\end{aligned}
\end{equation}
\par Equation (\ref{crossconstr}) is the main result of this section. We emphasize again that this is the irrational generalization of the crosscap constraint equation obtained previously in \cite{Fioravanti:1993hf,Pradisi:1995qy,Pradisi:1995pp} for the case of rational CFTs. The new ingredient here is the implementation of the Ponsot-Teschner kernel which allows us to make a statement about a generic two-dimensional CFT with $c>1$. It is important at this point to stress that equation (\ref{creq}) is an equality between \textit{distributions} on the Liouville momenta $P$, rather than an equality between usual functions. Indeed, the density $c_{12}(P)$ is a sum of delta functions and hence it only converges when integrated against some test function. This is an important point to keep in mind for our next discussion on the asymptotics. We will make use of some particular analytic properties of that kernel in the next section to derive a universal expression for the averaged Light-Light-Heavy data $C_{LLH}\Gamma_{H}$ encoded in (\ref{2ptdensity}).
\par Before we move on, let us highlight some other generic features of the crosscap constraint (\ref{crossconstr}). First, we observe that the constraint is \textit{linear} in the CFT data. This means that, assuming knowledge on the OPE coefficients $C_{12i}$ where $i$ is a scalar (and allowed by fusion rules to propagate in the OPE of $1$ and $2$), the equation allows us to determine only the \textit{ratios} of the $\Gamma_i$'s with some reference coefficient $\Gamma_{ref}$. Usually this reference is taken to be the identity, i.e. we end up calculating the ratios $\Gamma_i/\Gamma_{\mathbb{1}}$. Equivalently, given the spectrum of all the $\Gamma_i$'s we can use the crosscap constraint to instead evaluate the (ratios of) OPE coefficients of two operators $1,2$ that fuse into the scalar $i$. Furthermore, the spectral density $c_{12}(P)$ that encodes the CFT data is a distribution on a single variable $P$ (corresponding to an internal scalar operator with $h=\bar{h}$) and is, in general, \textit{not} positive definite. This is to be contrasted with the OPE-squared density that enters in the four-point function of primaries on the sphere. In that set up, the OPE-squared density is a distribution in two variables $P,\bar{P}$ for both the holomorphic and the anti-holomorphic sector, and one needs to encounter two copies of the fusion kernel to write down a constraint analogous to (\ref{crossconstr}) \cite{Collier:2018exn,Collier:2019weq}. In our present set up we had a single copy of the Virasoro block and therefore a single copy of the fusion kernel is needed to derive a constraint equation.

\subsection{Asymptotics from the fusion kernel}\label{sec:asymptOO}

We will now study the two-point function of identical external operators $O_1=O_2=O$ with dimensions $(h_O,\overline{h_O})$ and involution eigenvalue $\epsilon_O=\pm1$ on the crosscap. Equation (\ref{crossconstr}) in this case becomes
\begin{equation}\label{creqoo}
\begin{aligned}
c_{OO}(P_s)=\epsilon_O\int_C\frac{dP_t}{2}\mathbb{F}_{P_sP_t}\sbmatrix{P_O & P_O \\ \overline{P}_O & \overline{P}_O}c_{OO}(P_t)
\end{aligned}
\end{equation}
By considering identical external operators we allow the identity operator to propagate in both S and T-channels in (\ref{creq}). In particular, taking the kinematic limit $\eta\rightarrow1$ we see that the vaccum contribution dominates in the T-channel provided there is a \textit{gap} in the spectrum of scalars above the vacuum\footnote{The conformal blocks in (\ref{creq}) are normalized as $\mathcal{F}\sbmatrix{P_2 & P_1 \\ P_3 & P_4}(P|\eta)\sim\eta^{h_p-h_1-h_2}$ as $\eta\rightarrow0$.}. Therefore in the cross-channel, this should be compensated by an appropriate tail of heavy scalars $h_s\rightarrow\infty$, in much analogy with the usual lightcone bootstrap argument of the four-point function on the sphere\cite{Komargodski:2012ek,Fitzpatrick:2012yx}.
\par  Equivalently, we can avoid working with conformal blocks directly and instead start from (\ref{creqoo}) which can be written schematically as:
\begin{equation}\label{}
\begin{aligned}
c_{OO}(P_s)=\epsilon_O\Gamma_{\mathbb{1}}\mathbb{F}_{P_s\mathbb{1}}\sbmatrix{P_O & P_O \\ \overline{P}_O & \overline{P}_O}+\epsilon_O\sum_{\phi_t\in\{\text{scalars}\}}C_{OO\phi_t}\Gamma_t \mathbb{F}_{P_s P_{t}}\sbmatrix{P_O & P_O \\ \overline{P}_O & \overline{P}_O}
\end{aligned}
\end{equation}
In \cite{Collier:2018exn} it was shown that the non-vaccum kernels with T-channel dimension $h_t>0$ are \textit{suppressed} in the large $P_s\rightarrow\infty$ limit compared to the vacuum kernel as:\footnote{The result (\ref{fusionSuppression}) is accurate up to a factor independent of $P_s$, see e.g appendix B of \cite{Collier:2019weq} for more details. }
\begin{equation}\label{fusionSuppression}
\frac{\fusion_{P_s P_t}}{\fusion_{P_s\id}} \approx \begin{cases}
 	e^{-2\pi\alpha_t P_s} & \alpha_t = \tfrac{Q}{2}+iP_t \in (0,\tfrac{Q}{2}) \\
 	e^{-\pi Q P_s}\cos(2\pi P_t P_s) &  P_t \in \RR
 \end{cases}
 \quad\text{as }P_s\to\infty
\end{equation}
Assuming there is gap in the spectrum of scalar primaries $\phi_t$ above the identity, we immediately conclude that the cross-channel density should obey the asymptotics
\begin{equation}\label{AsFusion}
	c_{OO}(P_s)\sim \epsilon_O\Gamma_{\mathbb{1}}\mathbb{F}_{P_s\mathbb{1}}\sbmatrix{P_O & P_O \\ \overline{P}_O & \overline{P}_O} \ , \ \ \ \ \ \ \ P_s\rightarrow\infty
\end{equation}
Equation (\ref{AsFusion}) should be taken with a grain of salt. The spectral density $c_{OO}(P_s)$ is a sum of delta functions whereas the RHS is a smooth function of $P_s$. Therefore, and as we discussed earlier, the correct way to interpret (\ref{AsFusion}) is that the asymptotic relation holds when \textit{integrated} against some  appropriate test functions. The most conservative statement is that the asymptotic relation holds when we merely integrate over all states below some large Liouville momentum cutoff. We might also expect that (\ref{AsFusion}) would hold even when integrated over a small window around some high momentum $P_s$. The result however would depend in general on the size of that window, something that our asymptotic formula (\ref{AsFusion}) does not make explicit at all. A careful analysis of this sort implements the tools of Tauberian theory. Recent work towards this direction has been established in \cite{Qiao:2017xif,Mukhametzhanov:2018zja,Mukhametzhanov:2019pzy,Pal:2019yhz,Pal:2019zzr,Ganguly:2019ksp,Mukhametzhanov:2020swe,Das:2020uax} in the context of the Cardy formula and the various OPE coefficients asymptotics which arise in the orientable set up. It will be certainly interesting to apply analogous Tauberian theory methods in the various non-orientable asymptotic formulas that we obtain here, though we will not focus on that aspect in the present work.
\par With that in mind, we will now go ahead and analyze the RHS of (\ref{AsFusion}). The importance of the fusion kernel of the identity in the cross channel, namely $\mathbb{F}_{P_s\mathbb{1}}$, was first highlighted in \cite{Collier:2018exn} and later used in \cite{Collier:2019weq} to show that - along with its anti-holomorphic counterpart - it amazingly unifies the various heavy asymptotics of squared OPE coefficients in a single formula. We find yet another use of this kernel in (\ref{AsFusion}), namely governing the asymptotics of the Light-Light-Heavy product $C_{LLH}\times\Gamma_H$, except we now encounter only a \textit{single} copy of the kernel. Its analytic expression reads\cite{Collier:2018exn,Collier:2019weq}
\begin{equation}\label{eq:idFusion}
	\fusion_{P_s\id}\sbmatrix{P_2 & P_1 \\ P_2 & P_1} = \rho_0(P_s) C_0(P_1,P_2,P_s),
\end{equation}
where
\begin{equation}\label{eq:C0}
\begin{aligned}
         \rho_0(P) &= 4\sqrt{2}\sinh(2\pi bP)\sinh(2\pi b^{-1}P)\\
	C_0(P_1,P_2,P_3) &= \frac{1}{\sqrt{2}}{\Gamma_b(2Q)\over \Gamma_b(Q)^3}\frac{\prod_{\pm\pm\pm}\Gamma_b\left(\tfrac{Q}{2}\pm iP_1\pm iP_2 \pm iP_3\right)}{\prod_{k=1}^3\Gamma_b(Q+2iP_k)\Gamma_b(Q-2iP_k)}
	\end{aligned}
\end{equation}
The density $\rho_0(P)$ is related with the modular kernel of the identity torus character and essentially captures the (holomorphic half of the) asymptotics of the Cardy formula\cite{McGough:2013gka,Maxfield:2019hdt,Collier:2019weq}. We will meet this quantity again later when we discuss the Klein bottle (c.f. section \ref{sec:dualityKB}). For $C_0$, the $\prod$ in the numerator denotes the product of the eight combinations related by the reflections $P_k\to -P_k$. The function $\Gamma_b$ is a `double' gamma function, which is meromorphic, with no zeros, and with poles at argument $-mb-nb^{-1}$ for nonnegative integers $m,n$ (similarly to the usual gamma function, which has poles at nonpositive integers).  
\par In our present set up, we have
\begin{equation}\label{}
\begin{aligned}
\mathbb{F}_{P\mathbb{1}}\sbmatrix{P_O & P_O \\ \overline{P}_O & \overline{P}_O}=\rho_0(P) C_0(P_O,\overline{P}_O,P)
\end{aligned}
\end{equation}
and hence, the asymptotics at large (scalar) conformal dimension as $P\rightarrow\infty$ and fixed $P_O,\overline{P}_O$ gives:
\begin{equation}\label{cLLHGH}
\begin{aligned}
c_{OO}(P)\sim \epsilon_O\Gamma_{\mathbb{1}}2^{\frac{1}{2}-4P^2}e^{\pi Q P} P^{4(h_O+\overline{h_O})-{3Q^2+1\over 2}}{2^{Q^2-2\over 6}\Gamma_0(b)^6\Gamma_b(2Q)\over\Gamma_b(Q)^3\Gamma_b(Q+2iP_O)\Gamma_b(Q-2iP_O)\Gamma_b(Q+2i\bar{P}_O)\Gamma_b(Q-2i\bar{P}_O)}
\end{aligned}
\end{equation}
where $\Gamma_0(b)$ is a special function that appears in the large-argument asymptotics of $\Gamma_b$; see appendix A of \cite{Collier:2018exn} for more details. We can further make an asymptotic statement about the \textit{microcanonical average} of the product $C_{LLH}\times\Gamma_H$ by dividing with the asymptotic Cardy density of heavy scalar primaries which consists of two copies (holomorphic and anti-holomorphic) of $\rho_0(P)$, which for the case of scalars is just $\rho(P,\bar{P})\sim(\rho_0(P))^2$. This gives us\\
\begin{equation}\label{averageCLLHGH}
\begin{aligned}
\overline{c_{OO}(P)}&\sim \epsilon_O\Gamma_{\mathbb{1}}\rho_0^{-1}(P) C_0(P_O,\overline{P}_O,P)\\
&\sim  \epsilon_O\Gamma_{\mathbb{1}}2^{-\frac{1}{2}-4P^2}e^{-3\pi Q P} P^{4(h_O+\overline{h_O})-{3Q^2+1\over 2}}{2^{Q^2-2\over 6}\Gamma_0(b)^6\Gamma_b(2Q)\over\Gamma_b(Q)^3\Gamma_b(Q+2iP_O)\Gamma_b(Q-2iP_O)\Gamma_b(Q+2i\bar{P}_O)\Gamma_b(Q-2i\bar{P}_O)}
\end{aligned}
\end{equation}
This is our main universal asymptotic formula for the non-orientable CFT data, arising from the crosscap constraint of two-point functions of identical operators on $\mathbb{RP}^2$. Omitting order one coefficients and re-introducing the scaling dimensions $\Delta_i=h_i+\bar{h}_i$, we can rewrite (\ref{averageCLLHGH}) more compactly as
\begin{equation}\label{}
\begin{aligned}
\overline{C_{OOO_p}\Gamma_{O_p}}\approx 4^{-\Delta_p}e^{-3\pi\sqrt{\frac{c-1}{12}\Delta_p}}\Delta_p^{2\Delta_O-\frac{c+1}{8}}
\end{aligned}
\end{equation}
in the limit $\Delta_p>>c,\Delta_O$.

\section{CFT on $\mathbb{K}^2$}\label{sec:K2}

\subsection{Partition function}

The \textit{Klein bottle} $\mathbb{K}^2$ is a non-orientable surface with no boundary which can be thought of in many equivalent ways. Intuitively it can be constructed starting from the cylinder where space is compactified on a circle $\sigma\sim\sigma+2\pi$ and then compactify time $\tau\sim\tau+2\pi l$ but with the reverse orientation for $\sigma$. In other words (Fig.\ref{fig:KBex}),\
\begin{equation}\label{identif}
 (\tau,\sigma)\sim(\tau,\sigma+2\pi)\sim(\tau+2\pi l,-\sigma)
\end{equation}
The modulus $l\in(0,\infty)$ is a real parameter that characterizes topologically different Klein bottles. More formally the Klein bottle can be thought of as the $\mathbb{Z}_2$ quotient of a rectangular torus $\mathbb{K}^2=T^2/\mathbb{Z}_2$, or the connected sum of two $\mathbb{RP}^2$, i.e $\mathbb{K}^2=\mathbb{RP}^2\#\mathbb{RP}^2$. Its Euler character is $\chi_{\mathbb{K}^2}=0$.
\begin{figure}
	\centering
	\includegraphics[width=.3\textwidth]{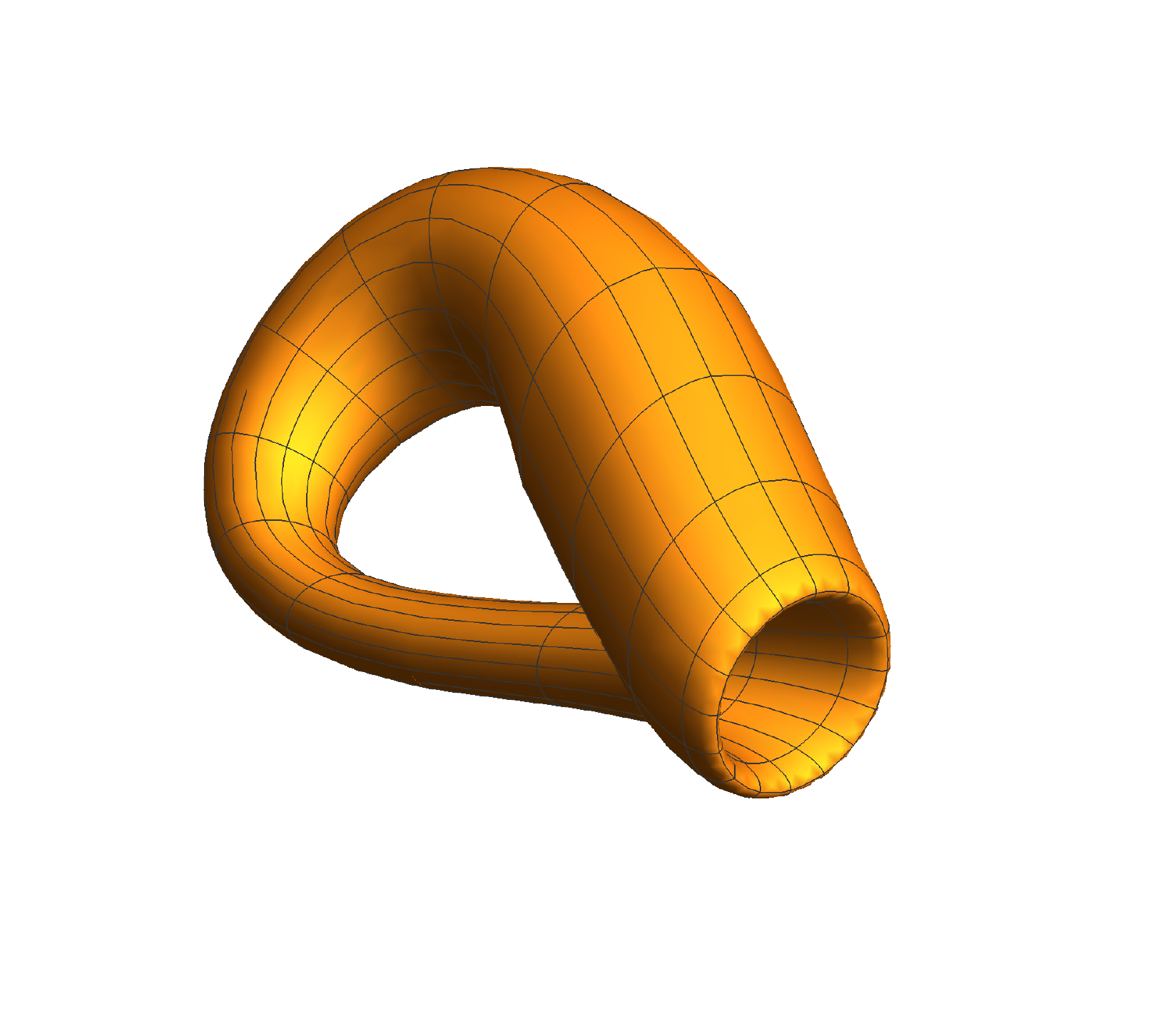} 
\caption[Klein bottle.]{The surface of the Klein bottle.
	\label{fig:KBex}}
	\end{figure}
\par We can easily construct the partition function of a CFT on the Klein bottle from a projection of the usual torus partition function of modulus $\tau=\frac{i\beta}{2\pi}$ ($\beta\in\mathbb{R}$), by inserting a \textit{parity} operator $\Omega$ at the final step of time evolution:
\begin{equation}\label{KBpf}
\begin{aligned}
Z_{\mathbb{K}^2}&=\frac{1}{2}\text{Tr}_{\mathcal{H}\times\overline{\mathcal{H}}}\left(\Omega q^{L_0-c/24}\bar{q}^{\bar{L}_0-c/24}\right)\\
&=\frac{1}{2}\sum_{i,j}\bra{i,\bar{j}}\Omega q^{L_0-c/24}\bar{q}^{\bar{L}_0-c/24}\ket{i,\bar{j}} \ \ \ \ \ \ \ \ \ \ , \ \ \ q=e^{2\pi i \tau}
\end{aligned}
\end{equation}
where $\ket{i,\bar{j}}$ denotes schematically an energy eigenstate in the total Hilbert space $\mathcal{H}_{tot.}=\mathcal{H}\times\overline{\mathcal{H}}$ on the circle, with $i$ labelling a state in the holomorphic sector $\mathcal{H}$ and $j$ a state in anti-holomorphic sector $\overline{\mathcal{H}}$. It's straight-forward to see how $\Omega$ acts on the state $\ket{i,\bar{j}}$\cite{Blumenhagen:2009zz}, namely 
\begin{equation}\label{parity}
\begin{aligned}
\Omega\ket{i,\bar{j}}=K_i\ket{j,\bar{i}} \ \ , \ \ \ \ \ \ \ \ \ \ \ \text{with} \ \  K_i=\pm1
\end{aligned}
\end{equation}
it interchanges the holomorphic and anti-holomorphic sectors, and the $K_i$'s are the eigenvalues of parity even/odd states. We note here that $K_i$'s are in general not necessarily equal to the eignevalues $\epsilon_i$ of the involution $\mathcal{I}$ that we defined back in (\ref{invo}), (\ref{invoprim}). Indeed, in cases where there is an extended symmetry algebra beyond Virasoro, the $K_i$'s and the $\epsilon_i$'s can be related by a phase factor \cite{Pradisi:1995pp,Stanev:2001na}. Since we are interested in CFT's with only Virasoro symmetry (and $c>1$) we're going to assume from now on that $K_i=\epsilon_i=\pm1$. 
\par Using that, we can massage the expression (\ref{KBpf}) into\footnote{We omit the conventional factor of $\frac{1}{2}$ in front of the partition function (\ref{KBpf}) from now on.}
\begin{equation}\label{massageKB}
\begin{aligned}
Z_{\mathbb{K}^2}(\beta)&=\sum_i\epsilon_i\bra{i,\bar{i}}\left(q\bar{q}\right)^{L_0-c/24}\ket{i,\bar{i}}\\
&=\sum_i\epsilon_i\bra{i,\bar{i}}e^{-2\beta(L_0-c/24)}\ket{i,\bar{i}}\\
&=\sum_{P\in\{\text{scalars}\}}\epsilon_P \chi_{P}\left(\frac{i\beta}{\pi}\right)
\end{aligned}
\end{equation}
In the first line, the effect of parity is to single out only those eigenstates in the trace which are symmetric under exchange of holomorphic and anti-holomorphic parts, i.e. the \textit{scalars}. This in turns leads to effectively identifying $L_0$ and $\bar{L}_0$ in the second line. Finally in the last line we organized the sum into \textit{scalar} Virasoro primary states $P$ with parity $\epsilon_P$, and $\chi_P(\tau)$ is the usual holomorphic character on the torus:
\begin{equation}\label{characters}
\begin{aligned}
\chi_{P}(\tau)&=\frac{e^{2\pi i \tau P^2}}{\eta(\tau)} \ \ , \ \ \ \ \ \ \ \ \ \ \ \ \ \ \ \ \ \ \ \ \ \ P\neq vac.\\
\chi_{vac.}(\tau)&=\frac{\left(1-e^{2\pi i \tau}\right)e^{-\frac{\pi i \tau Q^2}{2}}}{\eta(\tau)}
\end{aligned}
\end{equation}
In our normalizations, the relation between the Klein bottle modulus $l$ defined in (\ref{identif}) and $\beta$ in (\ref{massageKB}) is $l=\frac{\beta}{2\pi}$.
\par Before moving on to discussing the dual description of the Klein bottle, let's rewrite (\ref{massageKB}) in terms of even density distributions as
\begin{equation}\label{KBpfdens}
\begin{aligned}
Z_{\mathbb{K}^2}(\beta)&=\int_{-\infty}^{\infty}\frac{dP}{2}\rho_{sc.}^{\pm}(P)\chi_P(i\beta/\pi)\\
\end{aligned}
\end{equation}
where
\begin{equation}\label{}
\begin{aligned}
\rho_{sc.}^{\pm}(P)&\equiv \sum_{i\in\{\text{scalars}\}}\epsilon_i \left[\delta(P-P_i)+\delta(P+P_i)\right] \\
&=\rho_{\mathbb{1}}(P)+\sum_{i\neq\mathbb{1}}\epsilon_i \left[\delta(P-P_i)+\delta(P+P_i)\right]
\end{aligned}
\end{equation}
In the second line we distinguished the contribution of the identity which by assumption is parity invariant i.e. $\epsilon_{\mathbb{1}}=+1$, and also has degeneracy $n_{\mathbb{1}}=1$. The explicit expression reads
\begin{equation}\label{}
\begin{aligned}
\rho_{\mathbb{1}}(P)=\left[\delta\left(P-i\frac{b^{-1}+b}{2}\right)-\delta\left(P-i\frac{b^{-1}-b}{2}\right)+\left(P\leftrightarrow -P\right)\right]
\end{aligned}
\end{equation}
where the negative delta function is there to appropriately subtract the null state of the vacuum at level 1.
\par We're going to refer to the density $\rho_{sc.}^{\pm}(P)$ as the \textit{parity-weighted scalar density}.  A few comments about this quantity are in order. Unlike the usual spectral density defined from the torus partition function, the density $\rho_{sc.}^{\pm}(P)$ is supported only on scalar primaries $P$. Moreover it's \textit{not} positive definite because of the presence of the $\epsilon_i$'s: states with odd parity contribute negatively to the partition function (\ref{KBpfdens}). In the special case where $\epsilon_i=1$ for every scalar in the theory, the parity-weighted scalar density is equal to the \textit{total} scalar density:
\begin{equation}\label{}
\begin{aligned}
\left.\rho_{sc.}^{\pm}(P)\right|_{\epsilon_i=1} = \  \rho_{sc.}^{tot.}(P)\equiv\sum_{i\in\{\text{scalars}\}} \left[\delta(P-P_i)+\delta(P+P_i)\right]
\end{aligned}
\end{equation}
Our goal is to find an asymptotic expression for this density using the duality of the Klein bottle amplitude.

\subsection{Duality}\label{sec:dualityKB}

There is a dual way we can write the partition function on the Klein bottle  of modulus $\beta$; that is, if we think of the surface as a Euclidean evolution for time $\tau=\pi^2/\beta$ between two crosscap states $\ket{C}$. Then, the partition function is no longer a trace but rather a matrix element (or, a tree-level amplitude) between crosscap states\cite{Blumenhagen:2009zz}:
\begin{equation}\label{}
\begin{aligned}
\tilde{Z}_{\mathbb{K}^2}\left(\frac{\pi^2}{\beta}\right)&=\bra{ C}e^{-\frac{\pi^2}{\beta} H}\ket{C}\\
&=\bra{ C}e^{-\frac{\pi^2}{\beta} (L_0+\bar{L}_0-\frac{c}{12})}\ket{C}
\end{aligned}
\end{equation}
Expanding the crosscap state into Ishibashi states as in (\ref{ishi}) and using the normalization (\ref{normishi}) we get:
\begin{equation}\label{dualKB}
\begin{aligned}
\tilde{Z}_{\mathbb{K}^2}\left(\frac{\pi^2}{\beta}\right)&=\sum_{i,j}\Gamma_i\Gamma_j\delta_{ij}\langle\langle C_j|e^{-\frac{\pi^2}{\beta} (L_0+\bar{L}_0-\frac{c}{12})}| C_i \rangle \rangle\\
&=\sum_{P\in\{\text{scalars}\}}\Gamma_P^2\chi_P\left(\frac{i\pi}{\beta}\right)
\end{aligned}
\end{equation}
where  in the second line we again effectively identified the action of $L_0$ and $\bar{L}_0$ on the Ishibashi states. Therefore we ended up with a single holomorphic copy of a torus character with argument $\tau=-\frac{1}{i\beta/\pi}$. Notice that this is exactly the $\tau\rightarrow-1/\tau$ transform of the argument we had in (\ref{KBpfdens}). Using again an even density distribution we rewrite the above expression as:
\begin{equation}\label{KBpfdual}
\begin{aligned}
\tilde{Z}_{\mathbb{K}^2}\left(\frac{\pi^2}{\beta}\right)=\int_{-\infty}^{\infty} \frac{dP}{2} \rho_{\Gamma^2}(P)\chi_P(i\pi/\beta)
\end{aligned}
\end{equation}
where
\begin{equation}\label{}
\begin{aligned}
\rho_{\Gamma^2}(P)&\equiv \sum_{i\in\{\text{scalars}\}}\Gamma_i^2\left[\delta(P-P_i)+\delta(P+P_i)\right]
\end{aligned}
\end{equation}
We see that, unlike $\rho_{sc.}^{\pm}$, the density $\rho_{\Gamma^2}$ is manifestly positive definite in this dual description. In fact, it's obvious that the full amplitude defined by $\tilde{Z}_{\mathbb{K}^2}\left(\frac{\pi^2}{\beta}\right)$ in (\ref{dualKB}) is itself positive definite. This seems already a bit strange since $Z_{\mathbb{K}^2}(\beta)$ in (\ref{massageKB}) includes different signs for the parities and hence it's \textit{not} manifestly positive definite. We will comment more about this in section \ref{sec:bootstrap}.
\par Since the arguments of the characters $\chi_P$ in (\ref{KBpfdens}) and (\ref{KBpfdual}) are related with an $S-$transformation of the form $\tau\rightarrow-1/\tau$, we can always expand one character in a linear combination of the second by using the modular kernel\cite{McGough:2013gka, Maxfield:2019hdt,Collier:2019weq}: 
\begin{equation}\label{}
\begin{aligned}
\chi_P(\tau)&=\int_{-\infty}^{\infty}\frac{dP'}{2}\chi_{P'}(-1/\tau)\mathbb{S}_{P'P}
\end{aligned}
\end{equation}
\begin{equation}\label{modkern}
\begin{aligned}
\mathbb{S}_{P'P}&=2\sqrt{2}e^{-4\pi i  PP'}
\end{aligned}
\end{equation}
In other words, the modular S-transformation $\tau\rightarrow-1/\tau$ acts on individual characters $\chi_P$ as a Fourier transform in the Liouville parameter $P$. The only exception to this formula is the expansion of the vacuum representation, where we have to subtract the corresponding null state carefully. This is the \textit{identity S-matrix}  first introduced in \cite{McGough:2013gka}, and is equal to what we called $\rho_0(P)$ in section \ref{sec:asymptOO}:
\begin{equation}\label{idsm}
\begin{aligned}
\mathbb{S}_{P\mathbb{1}}&\equiv\mathbb{S}_{P,\frac{i}{2}\left(b+b^{-1}\right)}-\mathbb{S}_{P,\frac{i}{2}\left(b^{-1}-b\right)}=4\sqrt{2}\sinh{\left(2\pi bP\right)}\sinh{\left(2\pi b^{-1}P\right)}=\rho_0(P)
\end{aligned}
\end{equation}
Therefore, starting from (\ref{KBpfdens}) we can write
\begin{equation}\label{}
\begin{aligned}
Z_{\mathbb{K}^2}(\beta)&=\int_{-\infty}^{\infty}\frac{dP}{2}\rho_{sc.}^{\pm}(P)\chi_P(\tau) \ \ \ \ \ \ \ \ \ \  ,  \ \tau=i\beta/\pi\\
&=\int_{-\infty}^{\infty}\frac{dP}{2}\rho_{sc.}^{\pm}(P)\left[\int_{-\infty}^{\infty}\frac{dP'}{2}\chi_{P'}(-1/\tau)\mathbb{S}_{P'P}\right]\\
&=\int_{-\infty}^{\infty}\frac{dP'}{2}\left\{\int_{-\infty}^{\infty}\frac{dP}{2}\rho_{sc.}^{\pm}(P)\mathbb{S}_{P'P}\right\}\chi_{P'}(-1/\tau)
\end{aligned}
\end{equation}
Comparing with (\ref{KBpfdual}), the statement of equivalence of the partition functions on $\mathbb{K}^2$ translates to the corresponding densities into:
\begin{equation}\label{KBas1}
\begin{aligned}
\rho_{\Gamma^2}(P)&=\int_{-\infty}^{\infty}\frac{dP'}{2}\mathbb{S}_{P'P} \ \rho_{sc.}^{\pm}(P')\\
&=\hat{\rho}_{sc.}^{\pm}(P)
\end{aligned}
\end{equation}
This equation says that the Fourier transform of the parity-weighted scalar density, namely $\hat{\rho}_{sc.}^{\pm}(P)$, is equal to the density of the squared $\mathbb{RP}^2$ one-point function normalizations. Of course, since the densities we are interested in are \textit{even} and \textit{real}, we can also write the inverse Fourier transform as
\begin{equation}\label{KBas2}
\begin{aligned}
\rho_{sc.}^{\pm}(P)&=\int_{-\infty}^{\infty}\frac{dP'}{2}\mathbb{S}_{P'P} \ \rho_{\Gamma^2}(P')\\
&=\hat{\rho}_{\Gamma^2}(P)
\end{aligned}
\end{equation}
i.e. the \textit{same} kernel acts on $\rho_{\Gamma^2}(P)$ to give us $\rho_{sc.}^{\pm}(P)$\footnote{Because of this, we could equivalently express the modular kernel (\ref{modkern}) as $2\sqrt{2}\cos{\left(4PP'\right)}$ for our purposes.}.
\par The set of equations (\ref{KBas1}) and (\ref{KBas2}) is the main result of this section. They relate the non-orientable data of irrational CFTs for $c>1$ that is, the densities $\rho_{\Gamma^2}(P)$ and $\rho_{sc.}^{\pm}(P)$, by a single (holomorphic) copy of the usual modular kernel and hence they are 1d-Fourier transforms of each other\footnote{Note that since the distributions $\rho_{sc.}^{\pm}(P)$ and $\rho_{\Gamma^2}(P)$ can have support at imaginary values of $P$, i.e. from states with $h<\frac{c-1}{24}$, we should be careful with what we mean by their Fourier transform. As described in \cite{Maxfield:2019hdt} there is a solid mathematical background for these kind of distributions which can be taken to live on an enlarged space or, equivalently, a more restricted space of test functions which at least includes the characters $\chi_{P}$  (i.e the Gaussians). We refer to appendix A of \cite{Maxfield:2019hdt} for more details.}. Similar expressions have been written down before in the case of rational CFTs on the Klein bottle (see e.g. \cite{Blumenhagen:2009zz}), where the data $\Gamma_i$ and $\epsilon_i$ are related via (a single copy of) the usual finite dimensional $S$-matrix of the model. We see now that in the case of irrational CFTs we can make a similar statement by implementing the irrational version of the modular kernel (\ref{modkern}). As in the case of the crosscap constraint in section \ref{sec:RP2twopt}, we have to be careful to interpret the two equations as an equality between \textit{distributions} on the Liouville momenta $P$, rather than equality between usual functions. In the next section we will use the analytic properties of the modular kernel to derive asymptotic formulas for these densities.

\subsection{Free energy on $\mathbb{K}^2$ and asymptotics from the modular kernel}

One natural application of the the duality of the Klein bottle bottle partition function is to derive the free energy of the theory at high temperatures. We have
\begin{equation}\label{}
\begin{aligned}
Z_{\mathbb{K}^2}(\beta)=\tilde{Z}_{\mathbb{K}^2}\left(\frac{\pi^2}{\beta}\right)
\end{aligned}
\end{equation}
and hence for $\beta\rightarrow0$, the partition function is captured by the vacuum contribution in the dual channel given by (\ref{dualKB}):
\begin{equation}\label{ht}
\begin{aligned}
Z_{\mathbb{K}^2}(\beta)\sim\Gamma_{\mathbb{1}}^2e^{\frac{\pi^2c}{12\beta}}+\cdots \ ,  \ \ \ \ \ \ \ \ \beta\rightarrow0
\end{aligned}
\end{equation}
We then get
\begin{equation}\label{}
\begin{aligned}
\log{Z_{\mathbb{K}^2}(\beta)}\sim\frac{\pi^2c}{12\beta}+2\log\Gamma_{\mathbb{1}}+\cdots
\end{aligned}
\end{equation}
and from the standard thermodynamic formulas one can calculate
\begin{equation}\label{}
\begin{aligned}
\langle \Delta \rangle =-\partial_{\beta}\log{Z_{\mathbb{K}^2}(\beta)}&=\frac{\pi^2c}{12\beta^2}+\cdots\\
S=(1-\beta\partial_{\beta})\log{Z_{\mathbb{K}^2}(\beta)}&=\frac{\pi^2c}{6\beta}+2\log\Gamma_{\mathbb{1}}+\cdots \ ,  \ \ \ \ \ \ \ \ \beta\rightarrow0
\end{aligned}
\end{equation}
We see that the entropy includes a usual extensive contribution $\frac{\pi^2c}{6\beta}$ along with an explicit $o(1)$ contribution captured by $\Gamma_{\mathbb{1}}$. This is reminiscent of the entropy of a boundary CFT (see e.g. \cite{Cardy:2004hm}) where we get similar  $o(1)$ contributions depending on the corresponding boundary states that define the tree-level amplitude. In the case of the Klein bottle though we get a universal contribution which is just the logarithm of the partition function on $\mathbb{RP}^2$ (c.f. sect. \ref{sec:RP2onept}). On the opposite limit $\beta\rightarrow\infty$, the low temperature behaviour can be read from the direct channel (\ref{massageKB}):
\begin{equation}\label{lt}
\begin{aligned}
Z_{\mathbb{K}^2}(\beta)\sim \epsilon_{\mathbb{1}}e^{\frac{\beta c}{12}}+\cdots \ ,  \ \ \ \ \ \ \ \ \beta\rightarrow\infty
\end{aligned}
\end{equation}
where we kept explicit the dependence on the parity of the identity $\epsilon_{\mathbb{1}}$. The \textit{positivity} of the dual channel description of the Klein bottle $\tilde{Z}_{\mathbb{K}^2}$ now dictates that we can only have an \textit{even-parity} vacuum, namely $\epsilon_{\mathbb{1}}=+1$. Hence we see that the parity of the vacuum is basically fixed by the limit $\beta\rightarrow\infty$ of the Klein bottle duality.
\par In analogy with the derivation of the Cardy formula from the torus partition function\cite{Cardy:1986ie}, the limits (\ref{ht}),(\ref{lt}) imply some particular asymptotics for the cross-channel densities $\rho_{sc.}^{\pm}(P)$ and $\rho_{\Gamma^2}(P)$. To derive those asymptotics we will next work directly with the crossing relations (\ref{KBas1}) and (\ref{KBas2}) in the appropriate limits.
\par Let's first consider equation (\ref{KBas1}) in the limit of large $P$. Distinguishing the contribution from the identity we have the schematic sum
\begin{equation}\label{expG2}
\begin{aligned}
\rho_{\Gamma^2}(P)=\mathbb{S}_{P\mathbb{1}}+\sum_{i\in\{\text{scalars}\}}\epsilon_i \mathbb{S}_{P P_i}
\end{aligned}
\end{equation}
Assuming there is \textit{gap} in the spectrum of scalar primaries $P_i$ above the identity, the dominant contribution in the limit $P\rightarrow\infty$ comes from $\mathbb{S}_{P\mathbb{1}}$, because
\begin{equation}\label{}
\begin{aligned}
\frac{\mathbb{S}_{P P_i}}{\mathbb{S}_{P\mathbb{1}}}\sim\begin{cases}
&e^{-4\pi \alpha_iP} \ \ \ \ \ \ \ \ \ \ \ \ \ \ \ \ \ \ \ \ \ \ \ \ \   \ \ \ \ \alpha_i\in\left(0,\frac{Q}{2}\right) \ \ \  \text{discrete range}\\
&\cos{(4\pi PP_i)}e^{-2\pi QP} \ \ \ \ \ \ \ \ \ \ \ \ \ \ P_i\in\mathbb{R} \ \ \  \ \ \ \ \  \text{continuous range}
\end{cases}
\end{aligned}
\end{equation}
Therefore we find that the density of $\mathbb{RP}^2$ one-point function coefficients squared at large $P$ asymptotically approaches the identity S-matrix (\ref{idsm}), namely
\begin{equation}\label{gammaas}
\begin{aligned}
\rho_{\Gamma^2}(P)\sim\sqrt{2}e^{2\pi QP}\ \ \ \text{as} \ \ P\rightarrow\infty
\end{aligned}
\end{equation}
The microcanonical average of heavy $\mathbb{RP}^2$ one-point coefficients squared can be obtained, as before, after dividing with the Cardy density for heavy scalars  $\rho(P,\bar{P})\sim(\rho_0(P))^2$,
\begin{equation}\label{asavergeG}
\begin{aligned}
\overline{\rho_{\Gamma^2}(P)}\sim\rho^{-1}_0(P)\sim \frac{1}{\sqrt{2}}e^{-2\pi QP}\ \ \ \ \ \ \ \ \text{as} \ \ P\rightarrow\infty
\end{aligned}
\end{equation}
In terms of the scaling dimensions $\Delta=2h=2P^2+\frac{c-1}{12}$ and omitting order one terms, we can rewrite (\ref{asavergeG}) compactly as
\begin{equation}\label{avasG2}
\begin{aligned}
\overline{\Gamma^2_{\Delta}}\approx e^{-2\pi \sqrt{\frac{c-1}{12}\Delta}} \ , \ \ \ \ \ \ \ \text{for} \ \ \Delta>>c.
\end{aligned}
\end{equation}
This is the desired asymptotic formula arising from the duality of the Klein bottle partition function. Let us make some comments about (\ref{avasG2}). First we see that the average value of heavy $\Gamma^2$ is suppressed exponentially at large scalar dimensions. This came from the fact that we had to divide with \textit{two} factors of $\rho_0(P)$ - appropriately accounting for the Cardy density of heavy scalars - to get the average value from (\ref{gammaas}) which is given by a single factor of $\rho_0$. Secondly, it's worth stepping back and emphasizing that the IR data that basically determined the asymptotics of $\rho_{\Gamma^2}(P)$ via (\ref{expG2}) are \textit{universal}: it's the energy of the vacuum on the cylinder (or, the central charge) and the fact that the vacuum is parity-even $\epsilon_{\mathbb{1}}=+1$. In this sense, our asymptotic formula for the density is also universal for any 2d CFT consistently defined on the Klein bottle.  Of course, if one wishes to understand subleading contributions to this formula the result would depend on the dimensions and the parities of other light operators in the theory.
\par Finally, we can perform a similar analysis for (\ref{KBas2}) in the limit of large $P$. Distinguishing the contribution from the identity we have
\begin{equation}\label{}
\begin{aligned}
\rho_{sc.}^{\pm}(P)=\Gamma_{\mathbb{1}}^2\mathbb{S}_{P\mathbb{1}}+\sum_{i\in\{\text{scalars}\}}\Gamma_i^2\mathbb{S}_{PP_i}
\end{aligned}
\end{equation}
and hence, under the same assumptions of the existence of a gap in the scalar spectrum, we find that the weighted density of scalars at large $P$ approaches
\begin{equation}\label{}
\begin{aligned}
\rho_{sc.}^{\pm}(P)\sim \sqrt{2}\Gamma_{\mathbb{1}}^2e^{2\pi Q P} \ \ \ \text{as} \ \ P\rightarrow\infty 
\end{aligned}
\end{equation}
We see now that the heavy asymptotics of $\rho_{sc.}^{\pm}(P)$ is determined by \textit{different} light data, namely the dimension of the vacuum (as before) and the coefficient $\Gamma_{\mathbb{1}}$. Subleading corrections would require knowledge for further light operator dimensions and coefficients $\Gamma_i$. Rewriting in terms of scaling dimensions and omitting order one terms (but keeping $\Gamma_{\mathbb{1}}$ explicit) we obtain:
\begin{equation}\label{asrpm}
\begin{aligned}
\rho_{sc.}^{\pm}(\Delta)\approx \Gamma_{\mathbb{1}}^2 e^{2\pi\sqrt{\frac{c-1}{12}\Delta}} \ , \ \ \ \ \ \ \ \text{for} \ \ \Delta>>c.
\end{aligned}
\end{equation}
This result concludes our derivations of asymptotic formulas for the various non-orientable CFT data. We will next change gears and discuss a different implementation of the Klein bottle duality.

\subsection{Klein bottle bootstrap}\label{sec:bootstrap}

Let us now discuss a potential bootstrap application of the Klein bottle duality:
\begin{equation}\label{}
\begin{aligned}
Z_{\mathbb{K}^2}(\beta)=\tilde{Z}_{\mathbb{K}^2}\left(\frac{\pi^2}{\beta}\right)
\end{aligned}
\end{equation}
Expanding the two channels as in (\ref{KBpfdens}), (\ref{KBpfdual}) and bringing everything on one side, we end up with
\begin{equation}\label{KBbooteq}
\begin{aligned}
\int_{-\infty}^{\infty} \frac{dP}{4}\left[ \rho_{sc.}^{\pm}(P)\chi_P(i\beta/\pi)-\rho_{\Gamma^2}(P)\chi_P(i\pi/\beta)\right]=0
\end{aligned}
\end{equation}
Along the lines of the torus modular bootstrap program\cite{Hellerman:2009bu,Friedan:2013cba,Collier:2016cls,Afkhami-Jeddi:2019zci,Hartman:2019pcd}, we might want to call this the \textit{Klein bottle bootstrap equation}. It's an equation that should hold for any $\beta\in(0,\infty)$ for any consistent spectrum $\rho_{sc.}^{\pm}(P)$ (or equivalently $\rho_{\Gamma^2}(P)$, since they are Fourier transforms of each other) of a two-dimensional CFT on the Klein bottle with $c>1$ and just Virasoro symmetry. This equation relates a manifestly positive quantity, namely $\tilde{Z}_{\mathbb{K}^2}\left(\frac{\pi^2}{\beta}\right)$, with a not manifestly positive one, i.e. $Z_{\mathbb{K}^2}(\beta)$, so we already expect that it should be quite restrictive in terms of its allowed data.
\par For the torus modular bootstrap problem, a common line of attack is to to act on equations like (\ref{KBbooteq}) with an appropriate derivative functional and then evaluate the result at the self-dual temperature. For (\ref{KBbooteq}) the self-dual point is $\beta^*_{\mathbb{K}^2}=\pi$. Without considering any derivative functionals on $\beta$ let's just evaluate (\ref{KBbooteq}) at this self-dual point:
\begin{equation}\label{KBboot}
\begin{aligned}
\int_{-\infty}^{\infty} \frac{dP}{4}\chi_P(i)\left[ \rho_{sc.}^{\pm}(P)-\rho_{\Gamma^2}(P)\right]=0\\
\int_{-\infty}^{\infty} \frac{dP}{4}\chi_P(i)\left[ \rho_{sc.}^{\pm}(P)-\hat{\rho}_{sc.}^{\pm}(P)\right]=0\\
\int_{-\infty}^{\infty} \frac{dP}{4}\chi_P(i)\left[ \rho_{sc.}^{\pm}(P)-\int_{-\infty}^{\infty}\frac{dP'}{2}\mathbb{S}_{P'P} \ \rho_{sc.}^{\pm}(P')\right]=0
\end{aligned}
\end{equation}
We see that, unlike the analogous expression for the modular bootstrap of the torus partition function (e.g. the case of spinless modular bootstrap) evaluated at the self-dual temperature $\beta^*_{\mathbb{T}^2}=2\pi$, the brute-force evaluation of (\ref{KBbooteq}) at $\beta=\beta^*_{\mathbb{K}^2}$ yields a result that is \textit{not manifestly} zero. This is quite obvious since the two dual descriptions of the Klein bottle involve \textit{different} CFT data in their expansions, namely the parities $\epsilon_i$'s and the coefficients $\Gamma_i^2$, whereas for the torus partition function we have the same physical spectrum with the same degeneracies at both sides of the equation.
\par Assuming a discrete spectrum of scalar primaries on the Klein bottle, we can write the first line of (\ref{KBboot}) schematically as:
\begin{equation}
\begin{aligned}
\chi_{vac}(i)(1-\Gamma_{\mathbb{1}}^2)+\sum_{P\neq\mathbb{1}}n_P\chi_P\left(i\right)\left[\epsilon_P -\Gamma_P^2\right]=0
\end{aligned}
\end{equation}
where $n_P$ is the degeneracy of the scalar primary $P$ and we distinguished the contribution from the identity $\mathbb{1}$ which is parity-invariant ($\epsilon_{\mathbb{1}}=+1$) and has degeneracy $n_{\mathbb{1}}=1$. Using the expressions for the characters (\ref{characters}) we calculate
\begin{equation}\label{massKBboot}
\begin{aligned}
\left(e^{2 \pi }-1\right) e^{\frac{\pi}{12}   (c-25)}(1-\Gamma_{\mathbb{1}}^2)+\sum_{P\neq\mathbb{1}}n_Pe^{-2 \pi  P^2}\left[\epsilon_P -\Gamma_P^2\right]=0\\
1-\Gamma_{\mathbb{1}}^2+\sum_{P\neq\mathbb{1}}n_P\left(\frac{e^{-2 \pi ( P^2+\frac{c-25}{24})}}{e^{2 \pi }-1}\right)\left[\epsilon_P -\Gamma_P^2\right]=0\\
1-\Gamma_{\mathbb{1}}^2+\sum_{\Delta>0}n_{\Delta}\left(\frac{e^{- \pi ( \Delta-2)}}{e^{2 \pi }-1}\right)\left[\epsilon_\Delta -\Gamma_\Delta^2\right]=0
\end{aligned}
\end{equation}
where in the last line we switched variables from $P$ to $\Delta=2h=2P^2+\frac{c-1}{12}$ for the scalars. Notice that, curiously, the final expression does \textit{not} depend explicitly on the central charge since it got cancelled when we divide with $\chi_{vac}(i)\neq0$.
\par Albeit simple looking, equation (\ref{massKBboot}) is quite non-trivial since $\left[\epsilon_\Delta -\Gamma_\Delta^2\right]\neq0$ in general, and therefore it seems that an intricate cancellation must occur between the summands in order for the net result to be zero. Note also that the term $\left[\epsilon_\Delta -\Gamma_\Delta^2\right]$ can take two possible values for a given primary: if $\Delta$ is parity-even then $\left[\epsilon_\Delta -\Gamma_\Delta^2\right]=1-\Gamma_\Delta^2$, whereas if $\Delta$ is parity-odd then $\left[\epsilon_\Delta -\Gamma_\Delta^2\right]=-1$, since $\Gamma_{\Delta}$ vanishes for parity-odd states as we explained in section \ref{sec:RP2onept}.  
\par Our proposal is that equation (\ref{massKBboot}) should be satisfied for a given spectrum of $\{\Delta,n_{\Delta},\epsilon_\Delta=\pm1\}$ (or equivalently $\{\Delta,n_{\Delta},\Gamma^2_\Delta\}$) of scalar primaries for any consistent two-dimensional CFT on the Klein bottle with $c>1$ and just Virasoro symmetry. It will be interesting to further investigate this equation and ask questions akin to the bootstrap, namely: is there an upper bound on the first excited \textit{scalar} $\Delta_1$ above the identity so that (\ref{massKBboot}) is satisfied? Of course one can also implement the usual strategies and act with derivative functionals (or integral kernels) on the original equation (\ref{KBbooteq}) and evaluate the resulting expression at the self-dual point $\beta^*_{\mathbb{K}^2}=\pi$. We leave these aspects for future study.
\par As a confirmation, we check equation (\ref{massKBboot}) in the cases of the Ising and the tri-critical Ising model in Appendix \ref{appB} by implementing known data, and verify that it's non-trivially true. 

\section{Relation with gravity}\label{sec:GR}

The asymptotic formulas we derived in sections \ref{sec:RP2} and \ref{sec:K2} apply in the corresponding kinematic limits for any two-dimensional irrational CFT with $c>1$ appropriately defined on a non-orientable surface. It will be interesting to ask then what is the gravitational interpretation of such expressions in the case where the CFT is dual to a weakly coupled theory on asymptotically AdS spacetime with a non-orientable boundary. Recent works on holography in a non-orientable set up include \cite{Verlinde:2015qfa,Nakayama:2015mva,Nakayama:2016xvw, Maloney:2016gsg,Lewkowycz:2016ukf,LeFloch:2017lbt,Wang:2020jgh,Giombi:2020xah,Hogervorst:2017kbj}. In such theories, first we would expect our asymptotic formulas to hold on an \textit{extended} regime of kinematics which are fixed in the large $c$ limit \cite{Hartman:2014oaa,Heemskerk:2009pn,ElShowk:2011ag}. Furthermore, in the case of asymptotic expressions for CFT data like the OPE coefficients or the $\mathbb{RP}^2$ one-point function coefficients we would expect that semi-classical gravity computations will reproduce the \textit{averaged} asymptotic formulas after coarse-graining over some heavy microstates\cite{Kraus:2016nwo}. With these in mind, in this section we will do a first step towards this understanding by studying the formal large central charge limit of our asymptotic formulas, postponing the detailed derivation from the gravity side to an upcoming work\cite{3dgravityNonOrientable}.

\subsection{$C_{LLH}\times\Gamma_H$}

For the two-point function of identical operators $O$ on $\mathbb{RP}^2$ we derived an asymptotic formula for the averaged product $C_{OOP}\times\Gamma_P$, namely the OPE coefficients of the two operators $O$ fusing into a scalar $P$ and the $\mathbb{RP}^2$ one-point function coefficient of $P$ (c.f. \ref{sec:asymptOO}): 
\begin{equation}\label{eq:idFusion}
	\overline{c_{OO}(P)}\sim \epsilon_O\Gamma_{\mathbb{1}}\rho_0^{-1}(P) C_0(P_O,\overline{P}_O,P) \ , \ \ \ \ \ \ \ P\rightarrow\infty
\end{equation}
We will now take the conformal dimensions of $O$ to be held fixed in the large $c$ limit, and the dimension of the internal scalar operator $P$ to scale with the central charge $c$. In terms of their Liouville momentum this means
\begin{equation}\label{}
\begin{aligned}
P_O=i\left(\frac{Q}{2}-bh_o\right) \ , \ \ \ \ \overline{P}_O=i\left(\frac{Q}{2}-b\overline{h_o}\right) \ , \ \ \ \ P=b^{-1}p
\end{aligned}
\end{equation}
with $h_o,\overline{h_o},p$ fixed as $b\rightarrow0$. In gravitational terms, we could think of the product $C_{OOP}\times\Gamma_P$ as the amplitude of the process where a black hole microstate of zero spin propagates in the internal channel of the two-point function of a light probe $O$ in the AdS geometry with $\mathbb{RP}^2$ boundary\cite{Verlinde:2015qfa,Nakayama:2015mva,Nakayama:2016xvw, Maloney:2016gsg,Lewkowycz:2016ukf,Giombi:2020xah}. Equivalently, we can think of this process as the $\mathbb{Z}_2$ quotient of the $2\rightarrow2$ scattering of identical light probes $O$ on the boundary of Euclidean $AdS_3$ (i.e. $S^2$) which has a black hole as intermediate state. 
\par The various semiclassical limits of $\rho_0$ and $C_0$ have been studied extensively in \cite{Chang:2015qfa,Chang:2016ftb,Collier:2018exn,Collier:2019weq}. Using these results in our case (\ref{eq:idFusion}), we obtain
\begin{equation}\label{}
\begin{aligned}
\end{aligned}
\overline{c_{OO}(b^{-1}p)}\sim \frac{\epsilon_O\Gamma_{\mathbb{1}}}{2\sqrt{2}\sinh{(2\pi p)}} \ e^{\frac{1}{b^2}I^{(1)}+I^{(2)}\log{b}+I^{(3)}+\mathcal{O}(b^2)}
\end{equation}
where
\begin{equation}
\begin{aligned}
I^{(1)}&=-2\pi p -1+ 3(I(0)+I(1))+I(1\pm 2i p)-I(-{1\over 2}\pm i p)-2I({1\over 2}\pm i p)-I({3\over 2}\pm i p)-{1\over 2}\log(2\pi)\\
I^{(2)}&= 2-4(h_o+\overline{h_o})\\
I^{(3)}&={1\over 2}\log{\Gamma(1\pm 2ip)\over 4\pi\Gamma(2h_o)^2\Gamma(2\overline{h_o})^2}+(1-h_o-\overline{h_o})\log{\Gamma(-{1\over 2}\pm i p)\over \Gamma({3\over 2}\pm i p)}.
\end{aligned}
\end{equation}
with $I(x) = \int_{1/2}^xdz~\log\Gamma(z)$, and the $\pm$ symbols indicate that we have to take into account all possible permutations, e.g. $I(x\pm y) \equiv I(x+y) + I(x-y)$ and $\Gamma(x\pm y) \equiv \Gamma(x+y)\Gamma(x-y)$.
\par At this point we notice that the term $I^{(1)}$ is strictly negative for any $p\in\mathbb{R}$. This shows that the average value is still exponentially suppressed semi-classically with the particular 'action' $I^{(1)}$, and confirms the intuition that an intermediate black hole microstate is entropically suppressed in such process \cite{Giddings:2009gj,Fitzpatrick:2011hu}. Also, the explicit expression for $C_0$ allows us to extract the explcit subleading corrections at large central charge., namely the terms $I^{(2)},I^{(3)}$. It will be interesting to confirm those corrections from a gravity calculation.
\par Finally, let us comment briefly on the expectation of an extended regime of validity of our formula in holographic theories\footnote{We thank Henry Maxfield for raising this point.}. In section \ref{sec:asymptOO} we showed that the spectral density of $C_{OOP}\times\Gamma_P$ is basically captured by the (holomorphic half of the) fusion kernel of the identity $\mathbb{F}_{P_s\mathbb{1}}$, namely
\begin{equation}\label{AsFusion2}
	c_{OO}(P_s)\sim \epsilon_O\Gamma_{\mathbb{1}}\mathbb{F}_{P_s\mathbb{1}}\sbmatrix{P_O & P_O \\ \overline{P}_O & \overline{P}_O} \ , \ \ \ \ \ \ \ P_s\rightarrow\infty
\end{equation}
As we explained there, these kind of asymptotic equalities should really hold when \textit{integrated} against some window around the large momentum $P_s$, and hence knowledge of the \textit{support} of the fusion kernel $\mathbb{F}_{P_s\mathbb{1}}$ is important. It was highlighted in \cite{Collier:2018exn} that, if the external operators have sufficiently low twist, the support of $\mathbb{F}_{P_s\mathbb{1}}$ has a continuum for $h_s\geq\frac{c-1}{24}$ (or, $P_s\in\mathbb{R}$) but it also receives contributions from a discrete series of double-twist states with $h_s<\frac{c-1}{24}$ (or, imaginary $P_s$). The authors refereed to this spectrum (along with its OPE data) as the Virasoro Mean Field Theory (VMFT). In deriving (\ref{AsFusion2}) we didn't have to worry about those discrete double-twist states at all (even if we take our 'external' dimensions $h_O,\overline{h}_O$ to be sufficienlty low) since the asymptotic formula relies only on a \textit{single} limit, namely $h_s\rightarrow\infty$ for the internal scalar operator. However, in holographic theories one might expect that results like (\ref{AsFusion2}) would apply for an \textit{extended} regime of dimensions $h_s$ and therefore, the contributions from those discrete VMFT states might become important. In particular, in a non-orientable holographic CFT these discrete states (if contributing at all) should necessarily be scalars with even parity $\epsilon_i=+1$ and their non-orientable CFT data (namely the product $C_{ijk}\Gamma_k$) would be captured by the appropriate residue of the fusion kernel $\mathbb{F}_{P_s\mathbb{1}}$, along the lines of \cite{Collier:2018exn}. It will certainly be interesting to explore this possibility more rigorously in the gravity side.

\subsection{$\rho^{\pm}_{sc.}$ and $\Gamma^2_H$}

Next we turn to the asymptotic formulas that arise in the Klein bottle. In \cite{Maloney:2016gsg} the authors constructed explicitly the semi-classical saddles that contribute to the partition function of pure $AdS_3$ with a Klein bottle boundary. They found that there exist only \textit{two} leading saddle geometries at large $c$: $(i)$ the $\mathbb{Z}_2$ quotient of thermal AdS and $(ii)$ the $\mathbb{Z}_2$ quotient of the non-rotating BTZ black hole. This is to be contrasted with the infinite family of saddles in the case of the torus boundary\cite{Maloney:2007ud}.
\par We would like to restate the arguments of \cite{Maloney:2016gsg} adjusted to our conventions\footnote{The authors of \cite{Maloney:2016gsg} used different conventions in their set up for the Klein bottle compared to ours. In particular the inverse temperature $\beta_{\text{\cite{Maloney:2016gsg}}}$ used in section 3.1 of their paper is related to what we called $\beta$ in (\ref{massageKB}),(\ref{dualKB}) via $\beta_{\text{\cite{Maloney:2016gsg}}}\times\beta=2\pi^2$.}. The saddle which arises as the $\mathbb{Z}_2$ quotient of thermal AdS is a \textit{smooth} geometry with no singular points and it's basically the Euclidean continuation of the $\mathbb{RP}^2$ geon of \cite{Louko:1998hc}. Matching to what we called $\beta$ in (\ref{massageKB}),(\ref{dualKB}),  the classical action of this saddle is given by $\frac{1}{2}$ the action of thermal AdS considered at temperature $\tilde{\beta}=2\beta$, namely:
\begin{equation}\label{gravi}
Z^{(i)}_{grav}(\beta)\sim e^{\frac{1}{2}\times\frac{\tilde{\beta} c}{12}+O(c^0)}=e^{\frac{\beta c}{12}+O(c^0)}
\end{equation}
which exactly reproduces the low temperature behaviour of the Klein bottle partition function (\ref{lt}). On the other hand, the saddle that arises from the $\mathbb{Z}_2$ quotient of the non-rotating BTZ has point (i.e. zero-dimensional) singularities at two $\mathbb{Z}_2$ fixed points and hence it's \textit{not} smooth. Its action is given again by $\frac{1}{2}$ the action of the non-rotating BTZ at temperature $\tilde{\beta}=2\beta$, plus a potential contribution from the singularities $I_{sing.}$ which is independent of $\beta$:
\begin{equation}\label{gravii}
Z^{(ii)}_{grav}(\beta)\sim e^{\frac{1}{2}\times\frac{\pi^2 c}{3\tilde{\beta}}+I_{sing.}+O(c^0)}=e^{\frac{\pi^2 c}{12\beta}+I_{sing.}+O(c^0)}
\end{equation}
In \cite{Maloney:2016gsg} the authors emphasized that it is crucial to really include these singular saddles in the definition of the semi-classical gravitational path integral, in order to reproduce the CFT expectations from the Klein bottle partition function. And indeed, the result (\ref{gravii}) matches the high temperature behaviour of the Klein bottle partition function (\ref{ht}) with $I_{sing.}\equiv2\log{\Gamma_{\mathbb{1}}}$. One thing we would like to add at this point is that we expect in general $\Gamma_{\mathbb{1}}$ to be a function of the central charge. However, to the best of our knowledge, no such explicit dependence is known in the literature for irrational CFTs with $c>1$. Hence, we can't really say a priori to which order in $c$ (or $G_N$) this factor contributes in the free energy of (\ref{gravii}). It is an interesting question to try and determine a potentially universal $c$-dependence of $\Gamma_{\mathbb{1}}$ either from gravity or the CFT side for the irrational case $c>1$. On the CFT side, we saw that by studying crossing symmetry of two-point function on $\mathbb{RP}^2$ or the Klein bottle duality we can't really specify $\Gamma_{\mathbb{1}}$ without knowing anything else. It will be interesting to ask if this can happen by studying consistency conditions on other higher genus non-orientable surfaces or even surfaces with boundary, e.g. the M{\"o}bius strip for $c>1$.
\par In any case, the Euclidean actions of these two gravity saddles allow us to calculate their entropy at leading order in $c$. For the $\mathbb{Z}_{2}$ quotient of the non-rotating BTZ saddle we calculate the thermodynamic energy and entropy
\begin{equation}\label{}
\begin{aligned}
\langle E\rangle^{(ii)}&=-\partial_{\beta}\log{Z^{(ii)}_{grav}(\beta)}=\frac{c\pi^2}{12\beta^2}+\cdots\\
S^{(ii)}&=(1-\beta\partial_{\beta})\log{Z^{(ii)}_{grav}(\beta)}=\frac{c\pi^2}{6\beta}+I_{sing.}+\cdots
\end{aligned}
\end{equation}
Putting these together we have
\begin{equation}\label{}
\begin{aligned}
S^{(ii)}(\langle E\rangle)&=2\pi\sqrt{\frac{c}{12}\langle E \rangle}+I_{sing.}+\cdots
\end{aligned}
\end{equation}
and hence using standard arguments and translating to the microcanonical entropy, we get an asymptotic density of states
\begin{equation}\label{entrsadd}
\begin{aligned}
\rho^{(ii)}(E)&=e^{S^{(ii)}(\langle E\rangle)}\\
&=e^{2\pi\sqrt{\frac{c}{12}\langle E \rangle}+I_{sing.}} \ , \ \ \ \ \ \ \langle E\rangle \sim c \ , \ c\rightarrow\infty
\end{aligned}
\end{equation}
This density of states is exactly our asymptotic formula (\ref{asrpm}) for the parity-weighted scalar density $\rho_{sc.}^{\pm}$, with $I_{sing.}\equiv2\log{\Gamma_{\mathbb{1}}}$. Therefore, it looks like the entropy of the non-rotating BTZ quotient of \cite{Maloney:2016gsg} really counts scalar black hole microstates weighted by $\pm$ for even/odd parities. Crucially, the result depends on $I_{sing.}$ which, as we said before, must be in general a function of the central charge; it is still quite unclear though how $I_{sing.}$ is related with $\Gamma_{\mathbb{1}}$ from the gravity side. It is also obvious by now that our arguments are parallel to the derivation and matching of the Cardy formula with the entropy of the non-rotating BTZ black hole in the torus case\cite{Strominger:1997eq}. In particular we see that the asymptotic formulas (\ref{asrpm}) and (\ref{entrsadd}) apply crucially in different limits. This suggests that there must be an analogous extended regime of validity of (\ref{asrpm}), along the lines of \cite{Hartman:2014oaa}. We will investigate  more of these aspects in \cite{3dgravityNonOrientable}.
\par Finally, by redefining $\beta\rightarrow \frac{\pi^2}{\beta}$ we can perform the same analysis for the smooth quotient of thermal AdS. The effective density of states that reproduces the result (\ref{gravi}) is of course the same as in (\ref{entrsadd}), modulo the $I_{sing.}$ term:
\begin{equation}\label{}
\begin{aligned}
\rho^{(i)}(E)&=e^{2\pi\sqrt{\frac{c}{12}\langle E\rangle}} \ , \ \ \ \ \ \ \langle E\rangle \sim c \ , \ c\rightarrow\infty
\end{aligned}
\end{equation}
This density of states seems to capture the asymptotics of the density of squared $\mathbb{RP}^2$ one-point function coefficients $\rho_{\Gamma^2}(P)$ as in (\ref{gammaas}). While the relation with our asymptotic formulas from the CFT side is suggestive, it remains rather mysterious how this relation works exactly in the gravity side and it definitely deserves to be better understood.

\section*{Acknowledgements}
	I would like to thank Scott Collier, Keshav Dasgupta, Alex Maloney, Henry Maxfield and Tokiro Numasawa for helpful discussions and comments on the draft. This research is supported by the Simons Foundation Grant No. 385602 and the Natural Sciences and Engineering Research Council of Canada (NSERC), funding reference number SAPIN/00032-2015. 

\appendix
\section{Compactified Boson on $\mathbb{K}^2$}\label{appA}
In this Appendix we will study the compactified two-dimensional free boson $X(z,\bar{z})$ on a circle of radius $R$:
\begin{equation}\label{}
\begin{aligned}
X(z,\bar{z})\sim X(z,\bar{z})+ 2\pi R n \ \ \ \ \ \ \ \ \ n\in\mathbb{Z}
\end{aligned}
\end{equation}
where the worldsheet manifold is a Klein bottle $\mathbb{K}^2$.

\subsection{Partition Function}

From (\ref{massageKB}), the partition function on the Klein bottle is
\begin{equation}\label{FBpfKB}
\begin{aligned}
Z^{FB}_{\mathbb{K}^2}(\beta)=\sum_i\epsilon_i\bra{i,\bar{i}}q^{2(L_0-1/24)}\ket{i,\bar{i}} \ , \ \ \ \ \   \ \ \ \ \ q=e^{-\beta}
\end{aligned}
\end{equation}
The Hamiltonian of the free boson reads
\begin{equation}\label{}
\begin{aligned}
L_0=\frac{1}{2}j_0^2+\sum_{k\geq1}j_{-k}j_k
\end{aligned}
\end{equation}
where $j_n(\overline{j_n})$ are the usual $U(1)$ modes, which under the parity operator $\Omega$ transform as\cite{Blumenhagen:2009zz}:
\begin{equation}\label{Omegajn}
\begin{aligned}
\Omega j_n\Omega^{-1}=\overline{j_n} \ \ \ \ , \ \ \ \ \ \ \ \ \ \Omega \overline{j_n}\Omega^{-1}=j_n
\end{aligned}
\end{equation}
The states on the Hilbert space have the form:
\begin{equation}\label{}
\begin{aligned}
\ket{m,n;m_1,m_2,\cdots n_1,n_2,\cdots}=j_{-1}^{m_1}j_{-2}^{m_2}\cdots\overline{j_{-1}}^{n_1}\overline{j_{-2}}^{n_2}\cdots\ket{m,n} \ \ \ \ \ \ \ \ \ , \ \ m_i,n_i\in\mathbb{Z}_{\geq0}
\end{aligned}
\end{equation}
which are descendants of the highest weight vectors $\ket{m,n}$:
\begin{equation}\label{freebosstates}
\begin{aligned}
j_0\ket{m,n}&=\left(\frac{m}{R}+\frac{R n}{2}\right)\ket{m,n}\\
\overline{j_0}\ket{m,n}&=\left(\frac{m}{R}-\frac{R n}{2}\right)\ket{m,n}\ \ \ \ \ \ \ \ \  \ \ m,n\in\mathbb{Z}
\end{aligned}
\end{equation}
To determine the states $\ket{i,\bar{i}}$ that contribute in (\ref{FBpfKB}) we need to find all those highest weight vectors which are \textit{invariant} under $\Omega$. Using (\ref{Omegajn}) we can investigate the $j_0$ eigenvalue of the state $\Omega\ket{m,n}$:
\begin{equation}\label{}
\begin{aligned}
j_0\left(\Omega\ket{m,n}\right)&=\Omega \left(\Omega^{-1}j_0\Omega\right)\ket{m,n}=\Omega \overline{j_0}\ket{m,n}\\
&=\left(\frac{m}{R}-\frac{R n}{2}\right)\Omega\ket{m,n}
\end{aligned}
\end{equation}
Hence we see that
\begin{equation}\label{parm}
\begin{aligned}
\Omega\ket{m,n}=\ket{m,-n}
\end{aligned}
\end{equation}
and therefore only the states (\ref{freebosstates}) with $n=0$ can contribute in the Klein bottle trace. Moreover it's obvious from (\ref{parm}) that all of these states come with positive parity $\epsilon_i=+1$:
\begin{equation}\label{}
\begin{aligned}
\Omega\ket{m,0}=+\ket{m,0}
\end{aligned}
\end{equation}
Going back to the partition function (\ref{FBpfKB}) we can now calculate
\begin{equation}\label{}
\begin{aligned}
Z^{FB}_{\mathbb{K}^2}(\beta)&=\sum_i\epsilon_i\bra{i,\bar{i}} q^{2(L_0-1/24)} \ket{i,\bar{i}}\\
&=\sum_{m,m_1,m_2,\cdots}\bra{m;m_1,m_2,\cdots}q^{2(L_0-1/24)}\ket{m;m_1,m_2,\cdots}
\end{aligned}
\end{equation}
and using the fact that:
\begin{equation}\label{}
\begin{aligned}
L_0\ket{m;m_1,m_2,\cdots}=\left[\frac{1}{2}\frac{m^2}{R^2}+\sum_{k=1}^{\infty}km_k\right]\ket{m;m_1,m_2,\cdots}
\end{aligned}
\end{equation}
we arrive at
\begin{equation}\label{FBresult}
\begin{aligned}
Z^{FB}_{\mathbb{K}^2}(\beta;R)&=q^{-1/12}\sum_{m\in\mathbb{Z}}q^{\frac{m^2}{R^2}}\left(\sum_{m_1,m_2,\cdots}q^{2\sum_{k\geq1}km_k}\right)\\
&=q^{-1/12}\sum_{m\in\mathbb{Z}}q^{\frac{m^2}{R^2}}\prod_{k=1}^{\infty}\frac{1}{1-q^{2k}}\\
&=\frac{1}{\eta(\frac{i\beta}{\pi})}\sum_{m\in\mathbb{Z}}q^{\frac{m^2}{R^2}}\\
&=\frac{1}{\eta(\frac{i\beta}{\pi})}\theta_3\left(\frac{i\beta}{\pi R^2}\right)
\end{aligned}
\end{equation}
where we identified the Jacobi theta function $\theta_3(\tilde{\tau})=\sum_{m\in\mathbb{Z}}\tilde{q}^{\frac{m^2}{2}}$ with $\tilde{q}=e^{2\pi i \tilde{\tau}}$.

\subsection{T-duality}

If we replace $R\rightarrow\frac{2}{R}$ 
in our previous result, we get
\begin{equation}\label{tdKB}
\begin{aligned}
Z^{FB}_{\mathbb{K}^2}\left(\beta;\frac{2}{R}\right)&=\frac{1}{\eta(\frac{i\beta}{\pi})}\theta_3\left(\frac{i\beta R^2}{4\pi }\right)\neq Z^{FB}_{\mathbb{K}^2}(\beta;R)
\end{aligned}
\end{equation}
So it appears that applying a naive T-duality transformation does not render the result invariant, contrary to what we might have expected. In fact, we can see immediately that the result (\ref{tdKB}) is equivalent to calculating a Klein bottle partition function for the free boson where only states (\ref{freebosstates}) with $m=0$
contributing in the trace. And this is indeed the case at the end, if we carefully take into account the effect of T-duality on the parity $\Omega$. In particular, $\Omega$ transforms under T-duality as \cite{Dabholkar:1997zd}:
\begin{equation}\label{}
\begin{aligned}
T\Omega T^{-1}=I \Omega
\end{aligned}
\end{equation}
where $I$ is an inversion that simply adds a minus sign on the Left-Right components of the free boson field $X(z,\bar{z})$ (see section 2.5 of \cite{Dabholkar:1997zd} for more details). This means that our previous transformation (\ref{Omegajn}) for the modes should be replaced under $T$-duality with
\begin{equation}\label{}
\begin{aligned}
(I\Omega)j_n (I\Omega)^{-1}=-\overline{j_n} \ \ \ \ , \ \ \ \ \ \ \ \ \ (I\Omega)\overline{j_n} (I\Omega)^{-1}=-j_n
\end{aligned}
\end{equation}
The only states that are invariant under the combined $I\Omega$ action have now the form:
\begin{equation}\label{}
\begin{aligned}
j_0\ket{0,n}&=\left(\frac{R n}{2}\right)\ket{0,n}\\
\overline{j_0}\ket{0,n}&=\left(-\frac{R n}{2}\right)\ket{0,n}\ \ \ \ \ \ \ \ \  \ \ m,n\in\mathbb{Z}
\end{aligned}
\end{equation}
i.e. they are indeed states (\ref{freebosstates}) with $m=0$. Hence, the result we get for the T-dual partition function is precisely (\ref{FBresult}) with $R\rightarrow\frac{2}{R}$.

\subsection{$\Gamma_{\mathbb{1}}^2$ for the free boson}

As a small application of the above calculation we can use the analytic expression (\ref{FBresult}) to compute the one-point function normalization $\Gamma_{\mathbb{1}}^2$ for the identity on $\mathbb{RP}^2$. We will do that by using the duality of the Klein bottle in the small $\beta$ limit.\\
\\
From the modular transformations of $\eta(\tau)$ and $\theta_3(\tau)$ we have:
\begin{equation}\label{}
\begin{aligned}
\theta_3(\tau)=\frac{\theta_3(-1/\tau)}{\sqrt{-i\tau}}\\
\eta(\tau)=\frac{\eta(-1/\tau)}{\sqrt{-i\tau}}
\end{aligned}
\end{equation}
Therefore, (\ref{FBresult}) can be written equivalently as
\begin{equation}\label{}
\begin{aligned}
Z^{FB}_{\mathbb{K}^2}(\beta)=R\frac{\theta_3(\frac{i\pi R^2}{\beta})}{\eta(\frac{i\pi}{\beta})}
\end{aligned}
\end{equation}
Expanding this as $\beta\rightarrow0$, we get
\begin{equation}\label{}
\begin{aligned}
Z^{FB}_{\mathbb{K}^2}(\beta)=Re^{\frac{\pi^2}{12\beta}}+\cdots
\end{aligned}
\end{equation}
And we notice that this has exactly the same form as the leading (vacuum) contribution of the dual channel Klein bottle partition function (c.f. (\ref{ht})):
\begin{equation}\label{}
\begin{aligned}
Z^{dual}(\beta)&=\Gamma_{\mathbb{1}}^2e^{2\pi i\left(\frac{i\pi}{\beta}\right)\left(-\frac{1}{24}\right)}+\cdots
\end{aligned}
\end{equation}
Hence, we find that the compactified free boson on the Klein bottle has
\begin{equation}\label{}
\begin{aligned}
\Gamma_{\mathbb{1}}^2=R.
\end{aligned}
\end{equation}

\section{Klein bottle bootstrap equation in Minimal Models}\label{appB}

In this Appendix we will investigate and verify the Klein bottle bootstrap equation (\ref{massKBboot}) in the case of two unitary Minimal Models, namely the Ising and the tri-critical Ising model.

\subsection{Ising model $\mathcal{M}(4,3)$}

The Ising model $\mathcal{M}(4,3)$ has three scalar primary operators $\phi_{(1,1)},\phi_{(2,2)},\phi_{(2,1)}$ with dimensions $h_{(1,1)}=0$, $h_{(2,2)}=\frac{1}{16}$ and $h_{(2,1)}=\frac{1}{2}$. In order to check (\ref{massKBboot})  we have to replace the characters $\chi_P(\tau=i)$ with the proper expressions for the degenerate characters $\chi_{r,s}(\tau=i)$ which for the $\mathcal{M}(p,p')$ minimal model read
\begin{equation}\label{}
\begin{aligned}
\chi^{(p,p')}_{r,s}(\tau)&=K^{(p,p')}_{r,s}(\tau)-K^{(p,p')}_{r,-s}(\tau)\\
K^{(p,p')}_{r,s}(\tau)&=\frac{1}{\eta(\tau)}\sum_{n\in\mathbb{Z}}q^{(2pp'n+pr-p's)^2/4pp'}
\end{aligned}
\end{equation}
with $(r,s)$ integers in the range $1\leq r\leq p'-1$ and $1\leq s\leq p-1$. The Klein bottle data for the Ising are known \cite{Fioravanti:1993hf}:
\begin{equation}\label{}
\begin{aligned}
&\epsilon_{(1,1)}=\epsilon_{(2,2)}=\epsilon_{(2,1)}=1\\
\Gamma_{(1,1)}=\sqrt{1+\frac{1}{\sqrt{2}}}& \ , \ \ \ \ \Gamma_{(2,2)}=0 \ , \ \ \ \ \Gamma_{(2,1)}=\sqrt{1-\frac{1}{\sqrt{2}}}
\end{aligned}
\end{equation}
Using that we can evaluate numerically the sum in (\ref{massKBboot}) for the Ising characters, namely
\begin{equation}\label{}
\begin{aligned}
\chi^{(4,3)}_{1,1}(i)\times\left(-\frac{1}{\sqrt{2}}\right)+\chi^{(4,3)}_{2,2}(i)+\chi^{(4,3)}_{2,1}(i)\times\left(+\frac{1}{\sqrt{2}}\right)\\
\end{aligned}
\end{equation}
and indeed check that it's zero.

\subsection{Tricritical Ising model $\mathcal{M}(5,4)$}

Similarly, for the tri-critical Ising model $\mathcal{M}(5,4)$  we have 6 scalar primaries 
\begin{equation}\label{}
\begin{aligned}
\phi_{(1,1)},\phi_{(1,2)},\phi_{(1,3)},\phi_{(1,4)},\phi_{(2,2)},\phi_{(2,4)}
\end{aligned}
\end{equation}
with dimensions
\begin{equation}\label{}
\begin{aligned}
h_{(1,1)}=0 \ , \ \ \ h_{(1,2)}=\frac{1}{10} \ , \ \ \ h_{(1,3)}=\frac{3}{5} \ , \ \ \ h_{(1,4)}=\frac{3}{2} \ , \ \ \ h_{(2,2)}=\frac{3}{80} \ , \ \ \ h_{(2,4)}=\frac{7}{16}
\end{aligned}
\end{equation}
The direct channel Klein bottle partition function for $\mathcal{M}(5,4)$ has been studied in \cite{Bianchi:1991rd} where, just like in the Ising case, we can take all the operators to have even parity:
\begin{equation}\label{}
\begin{aligned}
& \ \ \ \ \ \ \   \ \epsilon_{(1,1)}=\epsilon_{(1,2)}=\epsilon_{(1,3)}=\epsilon_{(1,4)}=\epsilon_{(2,2)}=\epsilon_{(2,4)}=1\\
\end{aligned}
\end{equation}
Using the $S$-matrix of the tri-critical Ising model we can compute the $\Gamma_i^2$'s via the finite-matrix form of our crossing relation (\ref{KBas1}), i.e.:
\begin{equation}\label{}
\begin{aligned}
\Gamma_i^2&=\sum_{j}S_{ij}\epsilon_j
\end{aligned}
\end{equation}
where the $S$-matrix for $\mathcal{M}(5,4)$ is a $6\times6$ matrix with components:
\begin{equation}\label{}
\begin{aligned}
S^{\mathcal{M}(5,4)}_{rs;\rho\sigma}&=\frac{2}{\sqrt{10}}(-1)^{1+s\rho+r\sigma}\sin{\left(\frac{4\pi}{5}r\rho\right)}\sin{\left(\frac{5\pi}{4}s\sigma\right)} 
\end{aligned}
\end{equation}
We find
\begin{equation}\label{}
\begin{aligned}
\Gamma^2_{(1,1)}=\left(1+\frac{1}{\sqrt{2}}\right) &\left(1+\frac{2}{\sqrt{5}}\right)^{1/2} , \  \Gamma^2_{(1,2)}=\left(1-\frac{1}{\sqrt{2}}\right) \left(1-\frac{2}{\sqrt{5}}\right)^{1/2} , \  \Gamma^2_{(1,3)}=\left(1+\frac{1}{\sqrt{2}}\right) \left(1-\frac{2}{\sqrt{5}}\right)^{1/2} , \\ 
&\Gamma^2_{(1,4)}=\left(1-\frac{1}{\sqrt{2}}\right) \left(1+\frac{2}{\sqrt{5}}\right)^{1/2}\ , \ \Gamma^2_{(2,2)}=\Gamma^2_{(2,4)}=0
\end{aligned}
\end{equation}
Using these data we evaluate numerically (\ref{massKBboot}) with the appropriate expressions for the characters of $\mathcal{M}(5,4)$, and verify again that the result is non-trivially zero.

\bibliographystyle{JHEP}
\bibliography{nonorientCFT}
\end{document}